\documentclass[useAMS,usenatbib]{mn2e}
\usepackage{color}
\usepackage{graphicx}
\usepackage{rotating}
\usepackage{lscape}

\title[Physical parameters of the SMC RR Lyrae stars]
{Physical parameters of the Small  Magellanic Cloud  RR Lyrae stars and the distance scale}
\author[Deb  \& Singh ]{Sukanta Deb$^1$\thanks{E-mail:
sdeb@physics.du.ac.in}, Harinder P. Singh$^{1,2}$ \\
$^{1}$Department of Physics \& Astrophysics, University of Delhi,
         Delhi 110007, India\\
$^{2}$CRAL-Observatoire de Lyon, CNRS UMR 142, 69561 Saint-Genis
Laval, France\\
}
\begin{document}

\date{Received on ; Accepted on }

\pagerange{\pageref{firstpage}--\pageref{lastpage}} \pubyear{2009}

\maketitle

\label{firstpage}

\begin{abstract}
We present a careful and detailed light curve analysis of RR Lyrae 
stars in the  Small Magellanic Cloud (SMC) discovered by the Optical
Gravitational Lensing Experiment (OGLE) project. Out of  536 single
mode RR Lyrae stars selected from the database, we have investigated
the physical properties of 335  `normal looking' RRab stars and 17
RRc stars that have good quality photometric light curves. We have
also been able to estimate the distance modulus of the cloud which
is in good agreement with those determined from other independent
methods. The Fourier  decomposition method has  been used to study the
basic properties of these variables. Accurate Fourier decomposition
parameters of 536 RR Lyrae stars in the OGLE-II database are
computed. Empirical relations between the Fourier parameters and
some physical parameters of these variables have been used to
estimate the physical parameters for the stars from the Fourier
analysis. Further, the Fourier decomposition of the light curves of
the SMC RR Lyrae stars yields their mean physical parameters as:
[Fe/H] = -1.56\,$\pm\,0.25$, M = 0.55
$\pm $\,0.01\,M$_{\odot}$, T$_{\rm eff} = 6404\, \pm 12$ K,
$\log \rm L = 1.60\,\pm0.01\,\rm L_\odot$ and M$\rm _V = 0.78\,\pm0.02 $ 
for 335 RRab variables and [Fe/H] = -1.90\,$\pm$\, 0.13, M = 0.82
$\pm $\,0.18\,M$_{\odot}$, T$_{\rm eff} = 7177\,\pm 16$ K, $\log\,\rm L = 1.62\,\pm \,0.02\,\rm L_{\odot}$ and M$\rm _V = 0.76\,\pm 0.05$ for 17 RRc
stars. Using the absolute magnitude together with the mean
magnitude, intensity-weighted mean magnitude and the phase-weighted
mean magnitude of the RR Lyrae stars, the mean distance modulus to
the SMC is estimated to be 18.86\,$\pm$0.01 mag, 18.83\,$\pm$0.01
mag and 18.84\,$\pm$0.01 mag respectively from the RRab stars.
From the RRc stars, the corresponding distance modulus values are found to be
18.92\,$\pm$0.04 mag, 18.89\,$\pm$0.04 mag and 18.89\,$\pm$0.04 mag
respectively. Since Fourier analysis is a very powerful tool for the
study of  the physical properties of the RR Lyrae stars, we
emphasize the importance of exploring the reliability of the
calculation of Fourier parameters together with the uncertainty
estimates keeping in view the large collections of photometric light
curves that will become available from variable star projects of the
future.
\end{abstract}

\section{Introduction}
RR Lyrae (RRL) stars have played an important role in determining
the cosmic distance scale in modern astronomy. While several studies
have used classical Cepheids as primary distance indicators, RRL
 stars have received substantial attention in solving the
distance scale problem.  Apart from distance determinations, RRL
 stars are of particular importance as a test bed for the
theories of stellar and galactic structure and evolution. They are
the tracers of old stellar populations of the bulge, disk and halo
components that are present everywhere.  RRL stars are radially
pulsating A-F variable stars with periods in the 0.2 - 1.2 day
range, and amplitude of variation $ \leq 2$ mag. They can be easily
identified, and play a key role as the cornerstone of the Population
II distance scale. They are extensively used to determine distances
to old and sufficiently metal-poor systems, where they are commonly
found in large numbers. In particular, RRL stars are present in
globular clusters (GCs) and the dwarf galaxies in the neighborhood
of the Milky Way (Greco et al. 2007), and have also been identified
in the M31 field (Brown et al. 2004, Dolphin et al. 2004), in some
M31 companions (Pritzl et al. 2005), and in  at least four M31 GCs
(Clementini et al. 2001). Distances to  the Large Magellanic Cloud (LMC) 
for the population II objects are based on the luminosity of the RRL stars 
(Clementini et al. 2002).

There have been a very few studies to estimate the distance scale of the SMC 
using the RRL stars. Using four RRL light curves of
the SMC cluster NGC 121, Walker \& Mack  (1988) have estimated the
distance modulus of the cluster to be 18.86$\pm$0.07 mag. 
On the  other hand, using 22 RR Lyrae stars surveyed around 1.3 square degrees
near the northeast arm of the SMC field NGC 361, Smith et al. (1992) obtained 
the distance modulus of the SMC as 18.90$\pm$0.16 mag. 
Also, there are other studies of estimating the 
distance scale of the SMC based on the binary star light curves and double 
mode Cepheids. Harries et al. (2003) reported that the distance modulus to the 
SMC is of the order of 18.89$\pm$0.14 mag taking 10 eclipsing binaries in the 
SMC. By selecting 40 eclipsing binaries of spectral type O 
and B in the SMC, Hilditch et al. (2005) have derived the fundamental 
parameters of the binaries and refined the distance modulus to the SMC to  
18.91$\pm$0.1 mag. Also Kov\'{a}cs et al. (2000) found the distance modulus of 
the SMC to be 19.05$\pm$0.017 mag based on the photometric data of double-mode
Cepheids from the OGLE project.

The stellar atmospheric parameters of effective temperature (T$_{\rm eff}$)
and surface gravity (log g) are of fundamental astrophysical
importance. They are the prerequisites to any detailed abundance
analysis and define the physical conditions in the stellar
atmosphere and hence are directly related to the physical properties
mass (M), radius (R) and luminosity (L) of the star. In this paper,
we present an independent analysis of 536 SMC RRL stars
discovered by the OGLE project (Soszy\'{n}sky et al. 2002). The OGLE
database is a very wealthy resource for studying the
characteristics of variable stars in the Galaxy, LMC and SMC. For the first time, we make use of the OGLE SMC
data to estimate the distance scale of the SMC using a large number
of well-sampled RRL light curves.  The road map of the present
investigation is to perform a Fourier analysis of the RRL stars
in order to estimate their physical parameters and hence the SMC
distance scale.

We employ the Fourier decomposition technique which is used
extensively to characterize the observed photometric light curves of
RRL and other types of variables. The accurate determination of
the Fourier coefficients is, therefore, an important task. We have
performed  an independent automated Fourier analysis of all the RRL
light curves selected in this paper by a computer code developed by
us. In section 2 we give a brief description of the database that we
use and the procedure of removing the outliers from the light curve
data. We present Fourier decomposition of the light curves in
section 3. We also describe the use of  the unit-lag
auto-correlation function for finding out the optimal order of the
fit to the RRL light curves. Section 4 describes the error
analysis of the Fourier decomposition parameters $\rm \phi_{i1}$ and
R$_{i1}$.  Section 5 describes the calibration of the I band  data
to the  V band. In section 6, we describe the various physical
parameters of the RRLs obtained by using empirical relations
from the literature. Section 7 describes the distance determination
of the SMC. Lastly, in section 8, we present the conclusions of our
study.

\section{THE OBSERVATIONAL DATABASE}
The RRL stars analyzed in the present work were selected from
the high quality photometric catalog of RRL stars in the SMC
from OGLE-II database (Soszy\' nski et al. 2002, hereafter SZ02).
The catalog contains B, V and I light curves of 58 RRc stars  and
478 RRab stars located in the 11 areas close to the bar of the SMC
in Johnson-Cousins photometric system. The target stars selected for
the present analysis were passed through a multi-pass nonlinear
fitting algorithm in IDL (Interactive Data Language) which is very efficient 
in removing the outliers
from the data set. Points lying $2\rm \sigma$ away from the fit are rejected
from the data set so that the objects with well sampled and accurate light
curves are available for the analysis. All the selected targets have
evenly covered I band light curves with about 100 - 400 data points
in I band with an internal accuracy of 0.03 - 0.13 mag in I band.
The B and V light curves in the SZ02 catalog are of lower quality
and have a limited number of data points (20-40). Therefore, we use
only the I band data for estimation of the Fourier parameters. 
In Fig.~1(a) and (c) we show the typical phased light curve of a RRab
and a RRc star in the database. The corresponding light curves with
outliers $\geq 2\rm \sigma$ removed are shown in panels (b) and (d)
respectively.
\begin{figure}
\centering
\includegraphics[height=8cm,width=8cm]{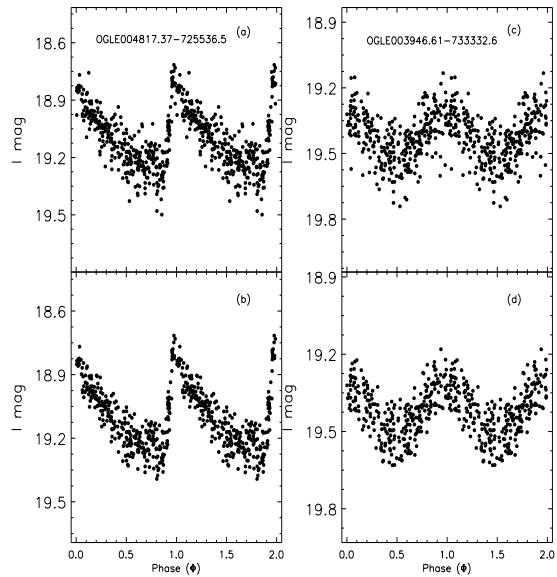}
\caption{Examples of light curve data from the OGLE-II database (a)
raw RRab light curve, (c) raw RRc light curve. In panels (b) and (d)
the corresponding outlier removed smoothed light curves are shown. }
\label{Fig 1}%
\end{figure}
\section{Fourier decomposition of the light curves}

\begin{figure}
\centering
\includegraphics[height=6cm,width=8cm]{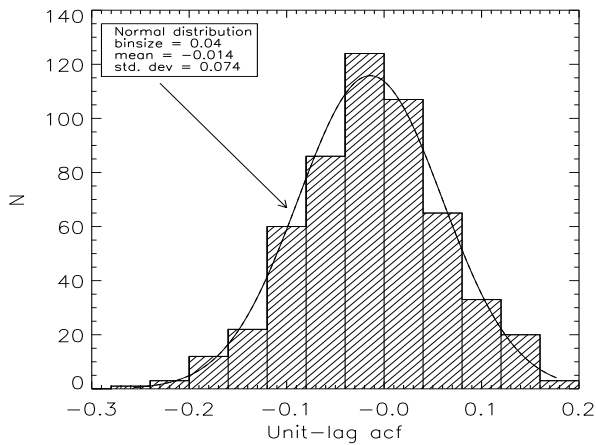}
\vspace{10pt}
\caption{Histogram plot of unit-lag auto-correlation function for 536 SMC
RRL stars in the OGLE database. The order of the fit is 4 and 5
respectively for the RRc and RRab stars.
 }
\label{Fig 2}%
\end{figure}
\begin{figure}
\centering
\includegraphics[height=6cm,width=8cm]{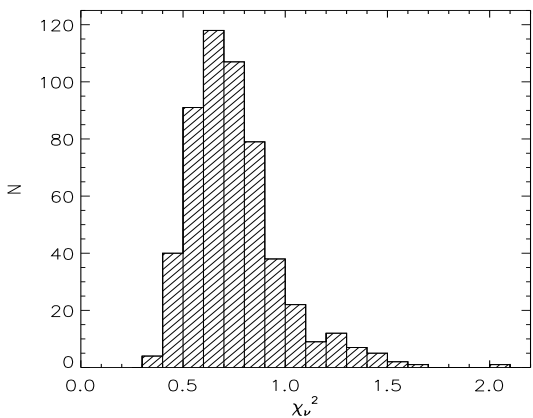}
\caption{Histogram plot of ${\rm \chi_{\rm \nu}}^{2}$ of 536 RRL stars selected
from the database.}
\label{Fig 3}%
\end{figure}
\begin{figure}
\centering
\includegraphics[height=12cm,width=8cm]{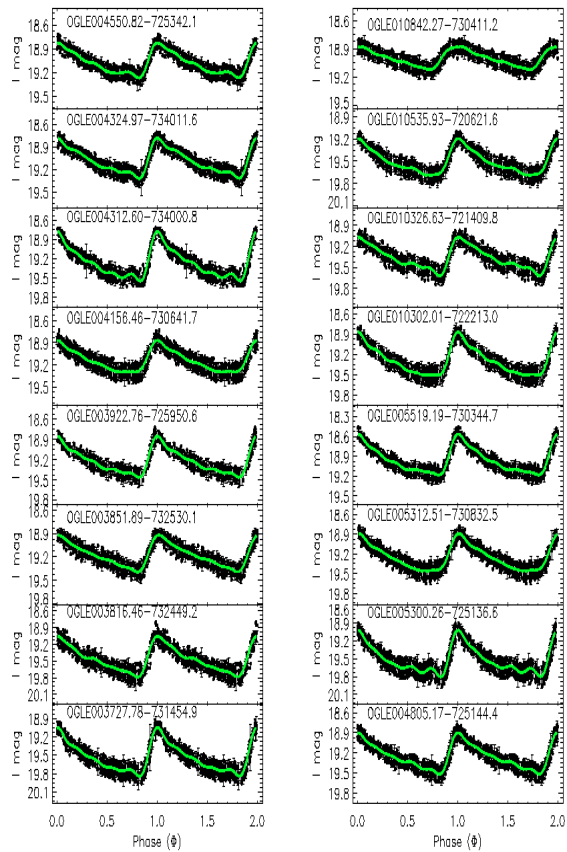}
\caption{Fourier fitted light curves of a selection of  RRab stars
after pre-processing of outliers removal. The order of the fit to the
light curve is 5. The solid lines show the Fourier fitted light
curves.}
\label{Fig 4}%
\end{figure}
\begin{figure}
\centering
\includegraphics[height=12cm,width=8cm]{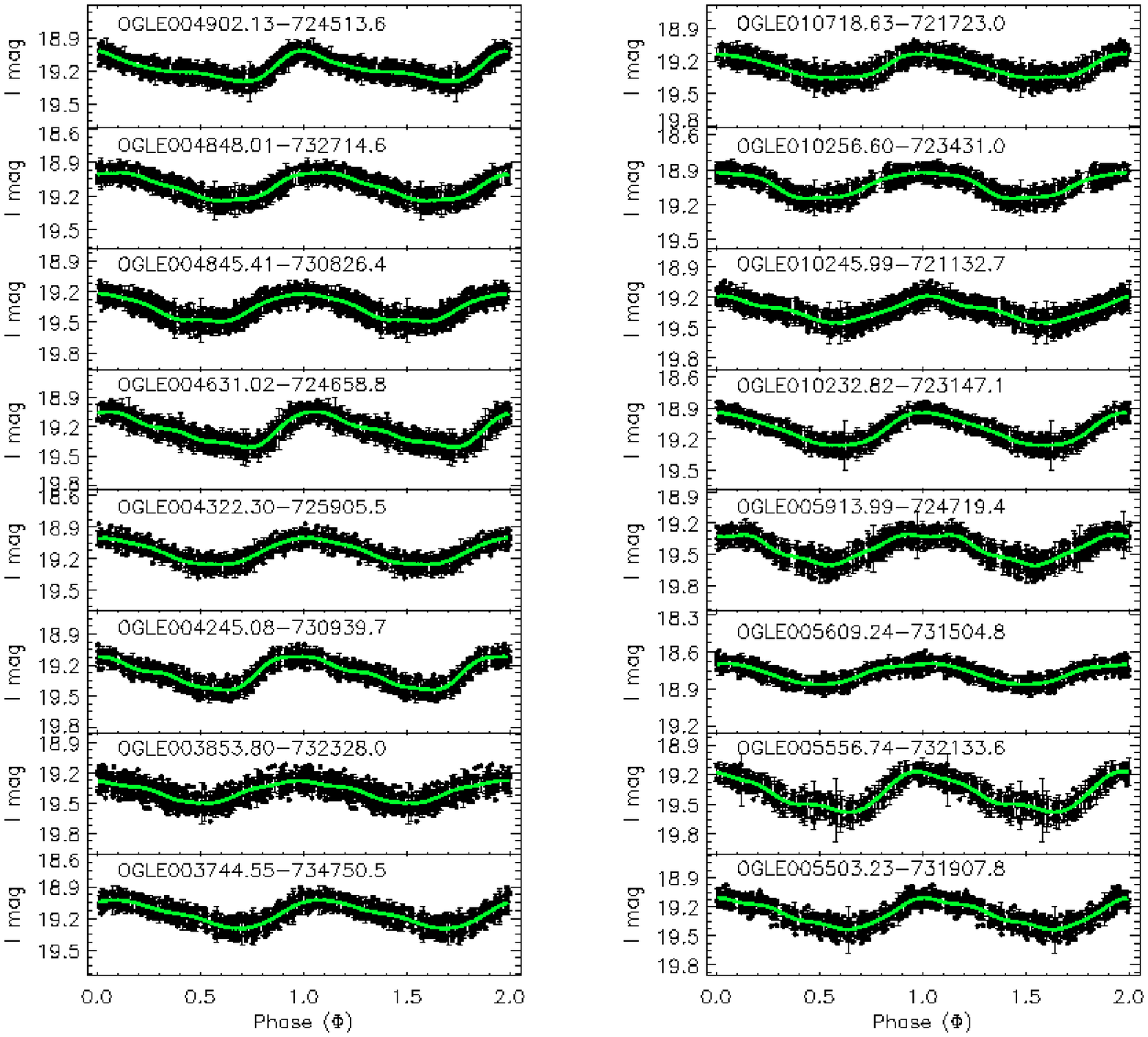}
\caption{Fourier fitted light curves of a selection of RRc stars
after removing the outliers. The order of the fit to the light curve
is 4. The solid lines are the Fourier fitted light curves.}
\label{Fig 5}%
\end{figure}

 Fourier
decomposition technique is a powerful  and robust technique for
describing the shape of the photometric light curves of the RRL and
other variable stars. The method has also  been utilized for
variable star classification (Deb \& Singh 2009), finding the
physical parameters of stars like absolute magnitude, metallicity,
effective temperature, gravity, helium abundance etc. of RRL
and other stars (Kov\'{a}cs \& Kupi 2007 and references therein).
The method in its modern form has been revived by Simon \& Lee
(1981) to describe the progression of Cepheid light curve with
increasing period called Hertzsprung progression. They showed that
the lower order Fourier parameters are able to completely describe
the progression of Cepheid light curves. With theoretical light
curves based on hydrodynamical models Simon \& Lee (1981), Simon \&
Clement (1993, hereafter SC93), Clement \& Rowe (2000) were able to
derive various physical parameters of stars. On the other hand
Kov\'{a}cs and his collaborators (Jurcsik \& Kov\'{a}cs 1996,
hereafter JK96, Kov\'{a}cs \& Jurcsik 1996, Jurcsik 1998, Kov\'{a}cs
1998 , Kov\'{a}cs \& Kanbur 1998, Kov\'{a}cs \& Walker 1999) were
able to estimate astrophysical parameters by establishing empirical
relationships between the Fourier parameters and the physical parameters of
stars in the field. A number of studies have made use of these 
empirical relations to derive the physical parameters of the RRL stars.
To list a few of these studies, Kaluzny et al. (2000) used these empirical
relations to estimate physical parameters of 26 RRab and 16 RRc
stars in M5, Pe\~{n}a et al. (2007) used these empirical relations
to estimate physical parameters of 7 RRL stars in Bootes and
Arellano et al. (2008) have determined parameters for few RRL
variables in NGC 5466. There are a number of other studies that have
made use of the Fourier coefficients to determine various parameters of 
astrophysical importance. Sandage (2004) studied the correlation of particular Fourier  
components of the light curves of RRab stars with metallicty discovered by Simon and  
later by Kov\'acs and his co-workers. 
On the other hand, Morgan et al. (2006,  2007) have derived empirical relations 
 for RRc stars connecting $[Fe/H]$, period (P) and Fourier parameter $\phi_{31}$. 
In all the above studies, it has been seen that Fourier parameters can indeed 
be linked with global stellar parameters such as luminosity, mass, temperature,
metallicty and  radius. Most of the physical parameter estimations of stars 
require accurate calculation of Fourier parameters and any inaccuracies in the
determination of Fourier parameters will reflect in the calculated physical 
parameters of the stars.

We computed Fourier decomposition coefficients for 536 RRL
stars in the SMC for the I band photometric data. The observed
magnitudes were fitted with a Fourier cosine series of the form
\begin{equation}
I(t) = A_{0}+\sum_{i=1}^{m} A_{i} \cos(i\omega (t-t_{0}) +  \phi_{ i}),
\end{equation} where $I(t)$ is the observed magnitude, $A_{0}$ is the mean magnitude, $\omega$=2$\pi/P$ is the angular frequency, $P$ is the period of the star
 in days, $t$ is the time of observation, $t_{0}$ is the epoch of maximum light, $A_{i}$ and $\phi_{i}$ are the $i$th order Fourier coefficients and $m$ is the order of the fit. Eqn.~(1) has $2m+1$
unknown parameters which require at least the same number of data
points to solve for these parameters.

Since period is known from the database, the observation time can be folded 
into phase
($\Phi$) as
\begin{displaymath}
\Phi =\frac{\left( t-t_{0}\right) }{P}-Int\left( \frac{\left(
t-t_{0}\right) }{P}\right).
\end{displaymath}
Here $t_{0}$ is the epoch of maximum light of the RRL light
curves. The value of $\Phi$ is from 0 to 1, corresponding to a full
cycle of pulsation and and $Int$ denotes the integer part of the quantity.
 Hence, Eqn.~(1) can be written as (Schaltenbrand
\& Tammann 1971)
\begin{equation}
I(t) = A_{0}+\sum_{i=1}^{m} A_{i}cos[2\pi i \Phi(t)+ \phi_{i}].
\end{equation}
The Fourier parameters are defined as
\[R_{i1}=\frac{A_{i}}{A_{1}}
\\; \, \phi_{i1}=\phi_{i}-i\phi_{1},\]
where $i > 1$. The $\phi_{i1}$ values have been adjusted to lie
between 0 and 2$\pi$ so that they are comparable to those available
in the literature.


The optimal order of the fit ($ m$) was
chosen to be 4 for RRc stars and 5 for RRab stars by the calculation
of unit-lag auto-correlation function (Fig.~2) and looking at the
$\rm \chi^{2}$ value of the fit. A detailed discussion on the optimal order of 
the fit has been given in Deb \& Singh (2009) which involves the calculation 
of unit-lag auto-correlation function using Baart's condition (Baart 1982, Petersen 1986). Increasing the order of the fit may reduce the $\chi^{2}$ to some 
extent but this will underestimate the distribution. The fits to the 
phase-folded light curves were made using Levenberg-Marquardt (LM) algorithm
which is based on $\chi^{2}$ minimization method (Press et al.
1992). The reduced chi-square for the fit to the RRL light
curves are nearly 1. The histogram plot of the ${\chi^{2}}_{\nu}$ of
the RRL  stars is shown in Fig.~3, where $\nu$ is the degree of freedom 
of the fit which is defined as the number of data points minus the number of 
parameters used to fit the light curve data. 

\begin{table*}
\caption{Fourier parameters for 478 fundamental mode (RRab) SMC RRL
variables in the OGLE database (I BAND DATA)}
\begin{center}
\scalebox{0.9}{%
\begin{tabular}{l cccccccccccccccc}
\\
\hline \hline ~~~~~~OGLE ID &Period&$\rm N_{\rm obs}$&${\rm \chi_{\rm
\nu}}^{2}$&$\rm \sigma_{\rm fit}$&$A_{\rm
I}$&$\rm A_{0}$&$\rm A_{1}$&$\rm R_{21}$&$\rm R_{31}$&$\rm R_{41}$&$\rm \phi_
{21}$&$\rm \phi_{31}$&$\rm \phi_{41}$&$\rm D_{m}$  \\
~~~~~~~~~~~(1)&(2)&(3)&(4)&(5)& (6)&(7)& (8)& (9)& (10)& (11)& (12)&(13)& (14)&(15) \\
\hline
\hline
004801.59-733021.5&  0.399572& 292& 0.503&     0.060&     0.821&    19.513&     0.268&  0.504&  0.357&  0.234&  4.105&  1.924&  0.187&  4.737 \\
005300.26-725136.6&  0.403584& 306& 0.716&     0.058&     0.786&    19.519&     0.257&  0.561&  0.311&  0.253&  3.785&  1.602&  5.713&  3.860 \\
003841.21-734422.9&  0.410569& 289& 0.664&     0.060&     0.843&    19.566&     0.285&  0.468&  0.344&  0.254&  3.888&  1.885&  6.001&  1.195 \\
003727.78-731454.9&  0.412698& 286& 0.815&     0.060&     0.799&    19.540&     0.276&  0.463&  0.336&  0.204&  4.116&  2.060&  0.093&  0.356 \\
005728.85-723454.6&  0.416258& 229& 0.966&     0.068&     0.765&    19.778&     0.258&  0.469&  0.362&  0.201&  4.344&  2.240&  0.636&  3.282 \\
005728.85-723454.6&  0.416260& 229& 0.966&     0.068&     0.765&    19.778&     0.258&  0.469&  0.362&  0.201&  4.344&  2.240&  0.636&  3.282 \\
005629.88-725213.0&  0.422334& 233& 0.708&     0.067&     0.258&    18.255&     0.098&  0.471&  0.190&  0.071&  4.310&  2.587&  6.114&  8.038 \\
005026.32-732418.2&  0.422681& 314& 1.024&     0.057&     0.630&    19.598&     0.222&  0.469&  0.338&  0.145&  4.164&  1.970&  6.021&  1.151 \\
004639.18-731324.7&  0.424320& 301& 0.501&     0.059&     0.873&    19.626&     0.306&  0.468&  0.324&  0.186&  4.053&  2.237&  0.076&  4.716 \\
003816.46-732449.2&  0.427897& 291& 1.273&     0.060&     0.734&    19.469&     0.262&  0.519&  0.279&  0.125&  4.367&  2.351&  0.986&  4.522 \\
005110.48-730750.0&  0.431720& 303& 0.715&     0.059&     0.879&    19.632&     0.316&  0.459&  0.303&  0.157&  4.249&  2.232&  0.095&  2.621 \\
010134.33-725427.4&  0.433496& 291& 1.001&     0.060&     0.645&    19.413&     0.234&  0.440&  0.235&  0.133&  4.165&  1.999&  5.184&  6.134 \\
010452.90-724025.9&  0.433623& 267& 0.502&     0.062&     0.580&    19.215&     0.192&  0.455&  0.384&  0.246&  3.910&  2.048&  6.086&  2.635 \\
005719.62-725540.7&  0.433862& 254& 0.627&     0.064&     0.747&    19.536&     0.256&  0.530&  0.283&  0.230&  4.080&  1.775&  0.061&  7.359 \\
005646.16-723452.2&  0.445986& 269& 0.762&     0.062&     0.774&    19.529&     0.264&  0.460&  0.339&  0.256&  3.823&  2.056&  0.169&  2.242 \\
005957.83-730647.6&  0.447303& 256& 0.686&     0.064&     0.784&    19.465&     0.277&  0.447&  0.297&  0.217&  3.913&  2.161&  0.240&  3.408 \\
005458.09-724948.9&  0.447471& 275& 0.976&     0.062&     0.663&    19.367&     0.239&  0.345&  0.352&  0.257&  4.128&  1.997&  0.022&  0.201 \\
004758.98-732241.3&  0.449794& 299& 0.520&     0.059&     0.764&    19.818&     0.253&  0.484&  0.333&  0.267&  4.059&  2.227&  0.090&  3.250 \\
010535.93-720621.6&  0.453841& 255& 0.699&     0.064&     0.498&    19.505&     0.191&  0.503&  0.179&  0.117&  4.094&  2.214&  6.035&  3.863 \\
010516.55-722526.5&  0.455958& 256& 0.599&     0.064&     0.756&    19.522&     0.255&  0.498&  0.315&  0.204&  4.267&  2.523&  0.391&  4.913 \\
\hline
\hline
\end{tabular}}
\end{center}
Complete table is available in the electronic form.
\end{table*}
\begin{table*}
\caption{Fourier parameters for 58 overtone mode (RRc) SMC RRL
variables in the OGLE database (I BAND DATA)}
\begin{center}
\scalebox{0.9}{%
\begin{tabular}{l ccccccccccccccc}
\\
\hline \hline OGLE ID &Period&$\rm N_{\rm obs}$&${\rm \chi_{\rm
\nu}}^{2}$&$\rm \sigma_{\rm fit}$&$A_{\rm
I}$&$\rm A_{0}$&$\rm A_{4}$&$\rm R_{21}$&$\rm R_{31}$&$\rm R_{41}$&$\rm \phi_
{21}$&$\rm \phi_{31}$&$\rm \phi_{41}$  \\
~~~~(1)&(2)&(3)&(4)&(5)& (6)&(7)& (8)& (9)& (10)& (11)& (12)&(13)&(14) \\
\hline
\hline
010220.67-723753.6& 0.261244& 281&     0.599&     0.061&     0.202&    19.611&     0.018&  0.066&  0.142&  0.193&  6.051&  3.361&  4.371\\
010246.69-725030.1& 0.262593& 293&     0.557&     0.060&     0.088&    18.801&     0.012&  0.453&  0.092&  0.319&  3.085&  4.728&  6.096\\
005609.24-731504.8& 0.277551& 265&     0.784&     0.063&     0.173&    18.774&     0.003&  0.103&  0.066&  0.032&  0.250&  4.089&  1.937\\
005451.72-723850.4& 0.277966& 270&     0.642&     0.062&     0.126&    19.027&     0.006&  0.348&  0.285&  0.115&  4.290&  3.298&  1.695\\
005115.64-724739.2& 0.280936& 313&     0.743&     0.058&     0.171&    19.161&     0.011&  0.192&  0.177&  0.128&  5.505&  3.101&  2.957\\
003908.31-735426.8& 0.281006& 297&     0.976&     0.059&     0.153&    18.985&     0.010&  0.306&  0.174&  0.163&  3.717&  5.884&  5.809\\
005601.43-724026.2& 0.283717& 271&     0.713&     0.062&     0.296&    19.584&     0.009&  0.363&  0.075&  0.078&  4.639&  5.659&  3.513\\
005754.46-723248.2& 0.283746& 253&     0.671&     0.064&     0.264&    18.520&     0.004&  0.190&  0.058&  0.033&  4.510&  5.230&  4.612\\
003706.77-730551.2& 0.293459& 292&     0.800&     0.060&     0.141&    19.198&     0.002&  0.236&  0.071&  0.030&  3.767&  2.998&  0.587\\
010245.99-721132.7& 0.294040& 300&     0.717&     0.059&     0.261&    19.326&     0.004&  0.145&  0.140&  0.032&  4.769&  0.719&  3.193\\
004744.89-725530.1& 0.297777& 306&     0.664&     0.058&     0.243&    19.185&     0.010&  0.311&  0.091&  0.088&  3.865&  0.441&  0.843\\
004633.73-725253.8& 0.302182& 293&     0.530&     0.060&     0.260&    19.501&     0.016&  0.123&  0.136&  0.135&  1.416&  4.446&  3.286\\
004631.02-724658.8& 0.308220& 300&     0.459&     0.059&     0.362&    19.241&     0.011&  0.322&  0.096&  0.071&  4.444&  2.469&  2.491\\
003946.61-733332.6& 0.309804& 273&     0.514&     0.062&     0.200&    19.420&     0.008&  0.046&  0.099&  0.080&  4.162&  1.948&  3.494\\
004902.13-724513.6& 0.310884& 314&     0.599&     0.057&     0.282&    19.180&     0.005&  0.541&  0.157&  0.053&  4.196&  2.315&  6.005\\
004245.08-730939.7& 0.311416& 286&     0.739&     0.060&     0.339&    19.277&     0.016&  0.297&  0.031&  0.104&  4.817&  3.105&  1.882\\
005503.23-731907.8& 0.313536& 278&     1.096&     0.061&     0.321&    19.275&     0.011&  0.171&  0.076&  0.075&  4.616&  2.970&  5.392\\
005913.99-724719.4& 0.313798& 252&     0.584&     0.064&     0.305&    19.440&     0.019&  0.078&  0.083&  0.128&  6.105&  3.581&  5.656\\
010249.92-720947.1& 0.316409& 277&     0.518&     0.061&     0.271&    19.379&     0.014&  0.173&  0.039&  0.110&  4.034&  1.202&  2.223\\
004330.74-733416.0& 0.320723& 289&     0.741&     0.060&     0.174&    19.293&     0.002&  0.096&  0.115&  0.021&  5.961&  2.734&  1.187\\

\hline
\hline
\end{tabular}}
\end{center}
Complete table is available in the electronic form.
\end{table*}

In Figs.~4 \& 5 we show smoothed (outliers removed) Fourier fitted
light curves for 16 randomly selected RRab and RRc stars
respectively. Further, the Fourier parameters for all the stars
computed from the I band data are given in Table 1 (for RRab) and
Table 2 (for RRc). $\rm N_{\rm obs}$ denotes the number of data points for
 each of the  light curves retained for the analysis after removing the
outliers. $\rm D_{m}$ is  the deviation parameter for RRab stars and is
discussed in Sec.~(6.1).
 In Fig.~6 we plot the  Fourier amplitude ratios $\rm R_{21}$, $\rm R_{31}$, $\rm R_{41}$ and phase differences $\rm \phi_{21}$, $\rm \phi_{31}$, $\rm \phi_{41}$ versus \rm $\log\rm P$ for the RRL dataset.
\begin{figure}
\centering
\includegraphics[height=8cm,width=8cm]{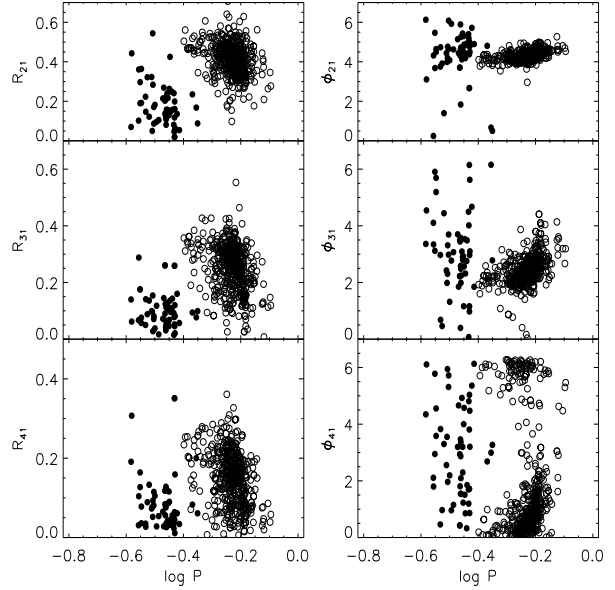}
\caption{Fourier phase differences  $\phi_{21}$ and  $\phi_{31}$  as
a function of log P  for the I band data of 536 program stars. Open
circles represent RRab stars and filled circles RRc stars.}
\label{Fig 6 }%
\end{figure}
\section{Errors in the Fourier parameters}

The error in $\rm x=\rm f(\rm u,v)$, where the standard errors in $\rm u's$ and
$\rm v's$ are known from LM method, is given by
\begin{equation}
{\rm \sigma}^{2}(\rm x)= {\rm \sigma_{\rm u}}^{2}{( \frac {\rm \partial {\it
x}}{\partial {\it u}})}^2+ {\rm \sigma_{\rm v}}^{2}( \frac {\partial
{\rm \it x}}{\partial {\rm \it v}})^2 +......+ 2 {\rm \sigma_{\rm uv}}^2(
\frac {\rm \partial {\rm \it x}}{\partial {\rm \it u}})( \frac {\rm \partial
\rm {\it x}}{\rm \partial {\rm \it v}}).
\end{equation}
In the case of a large number of observations Eqn.~(3) can be
reasonably approximated by
\begin{equation}
{\sigma}^{2}(x)= {\sigma_{\rm u}}^{2}{( \frac {\partial {\it x}}
{\partial {\it u}})}^2+ {\sigma_{\rm v}}^{2}( \frac {\partial {\it
x}}{\partial {\it v}})^2.
\end{equation}
Therefore the errors in  R$_{i1}$ and $\phi_{i1}$ can be
approximately written as
\begin{equation}
{\rm \sigma_{\rm R_{\rm i1}}}=\frac{1}{{\rm A_{1}}^2}\sqrt{(\rm
\sigma_{\rm A_{1}}^2{\rm A_{\rm i}}^2+\rm \sigma_{\rm A_{\rm
i}}^2{\rm A_{1}}^2)},
\end{equation}
\begin{equation}
{\rm \sigma_{\rm \phi_{\rm i1}}}=\sqrt{{\rm \sigma_{\rm \phi_{\rm
i}}}^2+\rm \rm i^{2}{\rm \sigma_{\rm \phi_{1}}}^2 }.
\end{equation}
In Figs.~7 \& 8 we plot the distribution of estimated errors in the
quantities R$_{21}$, R$_{31}$, R$_{41}$ and
 $\phi_{21}$, $\phi_{31}$ and $\phi_{41}$. As is quite obvious that the errors in
 $\phi_{31}$ are larger than the errors in
$\phi_{21}$ because the amplitudes for the higher order coefficients
are smaller and hence it is very difficult to derive their phase
with better precision. 

\begin{figure}
\centering
\includegraphics[height=8cm,width=8cm]{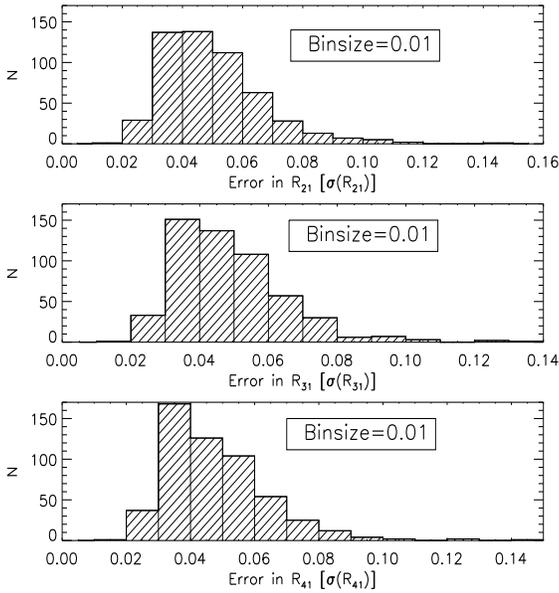}
\caption{Histogram plot of the distribution of  standard errors in R$_{21}$, R$_{31}$
and R$_{41}$ for the I band data. }
\label{Fig 7 }%
\end{figure}
\begin{figure}
\centering
\includegraphics[height=8cm,width=8cm]{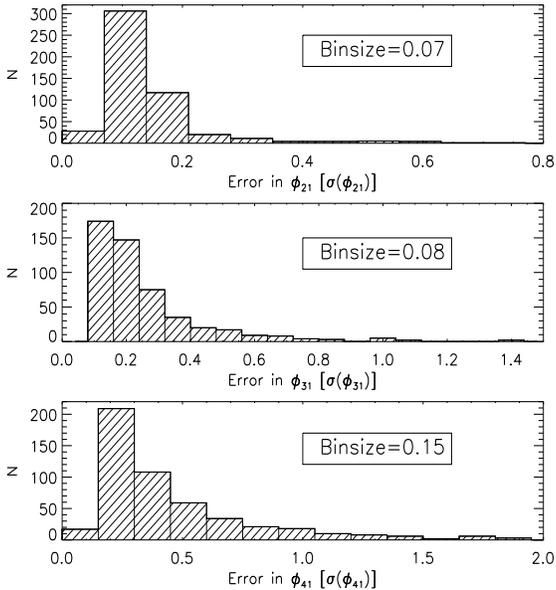}
\caption{Histogram plot of the distribution of  standard errors in $\phi_{21}$,
$\phi_{31}$ and $\phi_{41}$for the I band data. }
\label{Fig 8 }%
\end{figure}
\begin{figure}
\centering
\includegraphics[height=6cm,width=8cm]{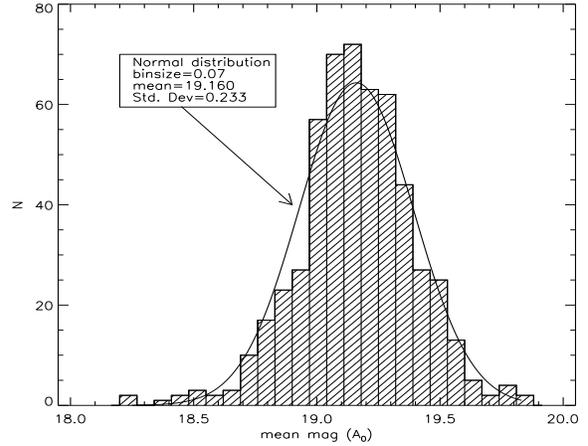}
\caption{Histogram plot of magnitudes of 536 RRL stars.
The solid line is the best fit normal distribution.}
\label{Fig 9}%
\end{figure}
\section{Calibration of the OGLE SMC RRL datasets }
\subsection{Fourier parameters in V band}

The Fourier coefficients of a particular light curve in 
different photometric bands are different. Nonlinear hydrodynamical models 
have been used by Dorfi \& Feuchtinger (1999, hereafter DF99) to obtain UBVI 
light curves of RRL stars. They have computed the inter-relations of the 
Fourier parameters in V and I bands and compared them with those obtained by 
Morgan et al. (1998, hereafter MSB98) from observed light curves of metal poor 
globular cluster M68 by Walker (1994, hereafter W94). However, only 8 RRab and 
8 RRc stars from W94 were used for setting up the relations in MSB98. While 
the V-I inter-relations for amplitude ratios ($R_{i1}$) of DF99 showed good 
agreement with the empirical relations of MSB98, large deviations showed up in 
the relations of some of the phase differences ($\phi_{i1}$) for two sets of 
models with Z = 0.001 and 0.0001. DF99 concluded that no distinct metallicity 
effect could be followed from these results and emphasized the need for more 
model calculations covering a wide range of stellar parameters.

In order to check the reliability of the theoretical inter-relations of the 
phase parameters of DF99, we have calculated the Fourier 
 phase parameters in the I and V bands from the highly accurate photometric 
light curves of selected RRab and RRc stars in globular cluster M3 from 
Benk\H{o} et al. (2006, hereafter B06) together with the data of 8 RRab and 8 
RRc stars from W94. All the RRab stars free from Blazhko 
effect and RRc stars having good quality light curves were selected from B06 
after visual inspection. We have used 29 RRab variables 1, 6, 9, 25, 27, 31, 
32, 42, 53, 57, 58, 69, 84, 89, 93, 94, 100,109,135,137, 139, 142, 144, 146, 
148, 165, 167, 175, 222 of globular cluster M3 from B06  and 8 variables 2, 9, 
10, 14, 22, 23, 25, 35 of globular cluster M68 from W94. In the case of RRc 
stars, 16 variables 21, 37, 75, 86, 88, 107, 126, 128, 131, 147, 152, 171, 177,
 208, 213, 259 of globular cluster M3 from B06 and 8 variables 1, 6, 11, 13, 
16, 18, 20, 24 of globular cluster M68 from W94 were used. We call this data 
set as B06+W94.  In Fig. 10, we show the V-I inter-relations of the Fourier 
phase parameters $\phi_{21}, \phi_{31}, \phi_{41}$ for RRab and RRc stars 
respectively. Points lying above a threshold sigma ($\sigma_{\rm thres}$) from 
the fit are rejected using multi-pass fitting algorithm in IDL . The threshold 
sigma level is chosen based on a reasonably high degree of correlation between
the parameters in the I and V band so that the number of data points for the 
fit retained are significant.  The mean difference of  0.61\% in absolute 
magnitude, 0.23\% in $\log\,(\rm L/L_{\odot})$, 4.03\% in [Fe/H], 0.10\% in 
effective temperature and 1.18\% in mass will result when DF99 inter-relations are used. 
\begin{figure*}
\centering
\includegraphics[height=14cm,width=18cm]{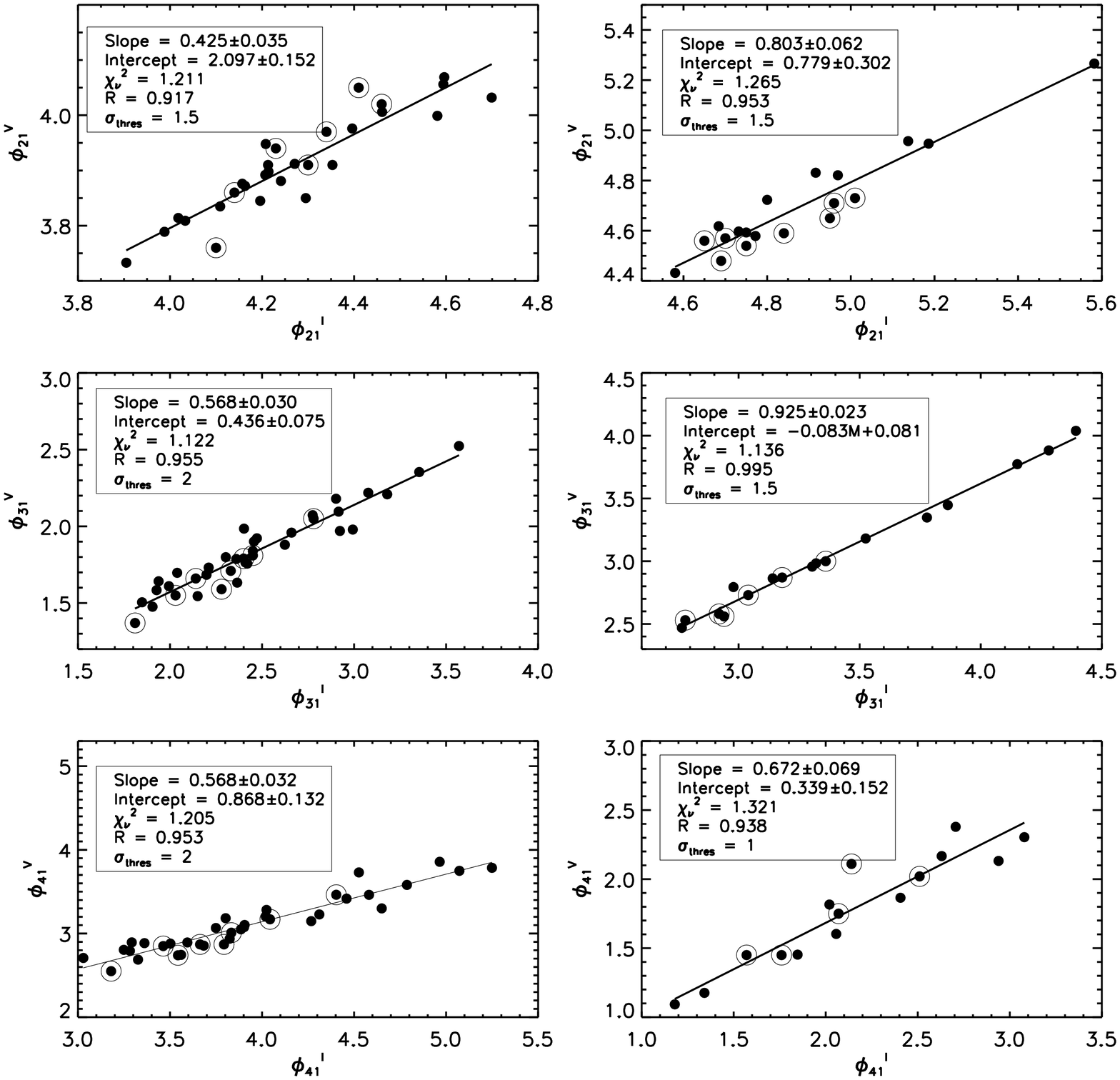}
\caption{Observational inter-relations between V and I band Fourier phase 
parameters. The left and the right panels show the inter-relations for the 
RRab and RRc stars respectively. The solid line is the linear fit to the 
data. The inter-relations are based on the cosine series fit. Dots represent 
the data points from B06 and encircled dots from W94 .
}
\label{Fig 10}%
\end{figure*}
\begin{figure*}
\centering
\includegraphics[height=14cm,width=18cm]{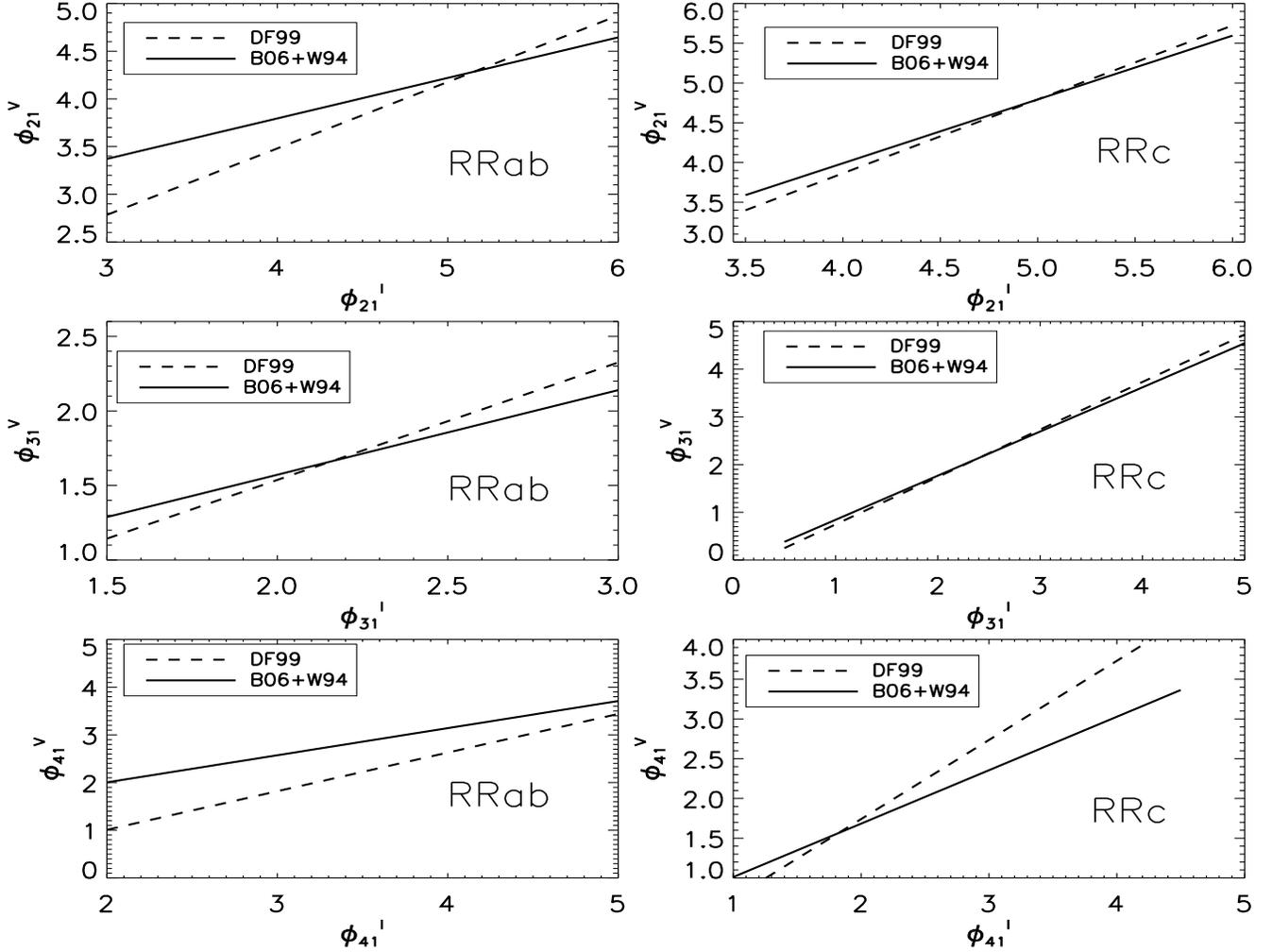}
\caption{Theoretical inter-relations of DF99 between V and I band Fourier
 phase parameters in comparison with the observational relations calculated 
from the combined data set of B06 and W94 (B06+W94). The inter-relations are 
based on the cosine series fit.}
\label{Fig 11}%
\end{figure*}
    
The Fourier inter-relations of the phase parameters from the I band
to the V band have the following form:
\begin{equation}
{\rm \phi_{i1}}^{\rm V} =\rm \alpha  +\rm \beta\,{\rm \phi_{i1}}^{\rm I}.
\end{equation}
\begin{table}
\caption{Coefficients of the Fourier inter-relations for
 RRab stars from DF99 and B06+W94 data. }
\begin{center}
\scalebox{0.9}{
\begin{tabular}{lccccc}
\hline
Relation &\multicolumn{2}{c}{DF99}& &\multicolumn{2}{c}{B06+W94} \\
\cline{2-3} \cline{5-6} \\
 & $\alpha$ & $\beta$ & & $\alpha$ & $\beta$ \\
 {$\phi_{21}$}$^{I}\rightarrow {\phi_{21}}^{V}$ & 0.693 & 0.697 & & 2.097\,$\pm$\,0.152&0.425\,$\pm$\,0.035 \\  
 {$\phi_{31}$}$^{I}\rightarrow {\phi_{31}}^{V}$ & -0.039 & 0.788 & &0.436\,$\pm$\,0.075 & 0.568\,$\pm$\,0.030 \\
 {$\phi_{41}$}$^{I}\rightarrow {\phi_{41}}^{V}$ & -0.611 & 0.810 & & 0.868\,$\pm$\,0.132 & 0.568\,$\pm$\,0.033 \\
\hline
\label{table 3}
\end{tabular}
}
\end{center}
\end{table}
\begin{table}
\caption{Coefficients of the Fourier inter-relations for
RRc stars from DF99 and B06+W94 data. }
\begin{center}
\scalebox{0.9}{
\begin{tabular}{lccccc}
\hline
Relation &\multicolumn{2}{c}{DF99}& &\multicolumn{2}{c}{B06+W94} \\
\cline{2-3} \cline{5-6} \\
 & $\alpha$ & $\beta$ & & $\alpha$ & $\beta$ \\
{$\phi_{21}$}$^{I}\rightarrow {\phi_{21}}^{V}$    & 0.144 & 0.930 & &0.779\,$\pm$ \,0.302 &0.803\,$\pm$\,0.062 \\  
 {$\phi_{31}$}$^{I}\rightarrow {\phi_{31}}^{V}$& -0.249 &0.995 & & -0.083\,
$\pm$\,0.081 &0.925\,$\pm$\,0.023 \\
 {$\phi_{41}$}$^{I}\rightarrow {\phi_{41}}^{V}$& -0.305 & 0.980 & & 0.339\,$\pm$\,0.153 & 0.672\,$\pm$\,0.069 \\
\hline
\label{table 3}
\end{tabular}
}
\end{center}
\end{table}

In Tables 3 and 4, we list the inter-relations obtained from the B06+W94 dataset
for RRab and RRc stars. Also given are the theoretical inter-relations of 
DF99. In Fig.~10, we plot the I and  V band Fourier parameters. In Fig.~11, 
we plot the theoretical inter-relations between the I and V band of DF99 and 
observational inter-relations obtained from the B06+W94 dataset.

To compute V band amplitude from I band amplitude we again resort to
the existing work in the literature. In Fig.~12 we plot V band
amplitudes as functions of I band amplitudes for RRL stars in 4
globular clusters, M68 from  Walker (1994), IC 4499 from Walker \&
Nemec (1996), NGC 1851 from Walker (1998) and M3  from Benk\H{o} et
al. (2006). Solid line denotes the linear relation as determined by a least
square fit given by
\begin{equation}
\rm A_{\rm V} = 0.071\,(\pm 0.019) + 1.500\,(\pm 0.040)\,\rm A_{\rm I}.
\end{equation}

\begin{figure}
\centering
\includegraphics[height=8cm,width=8.5cm]{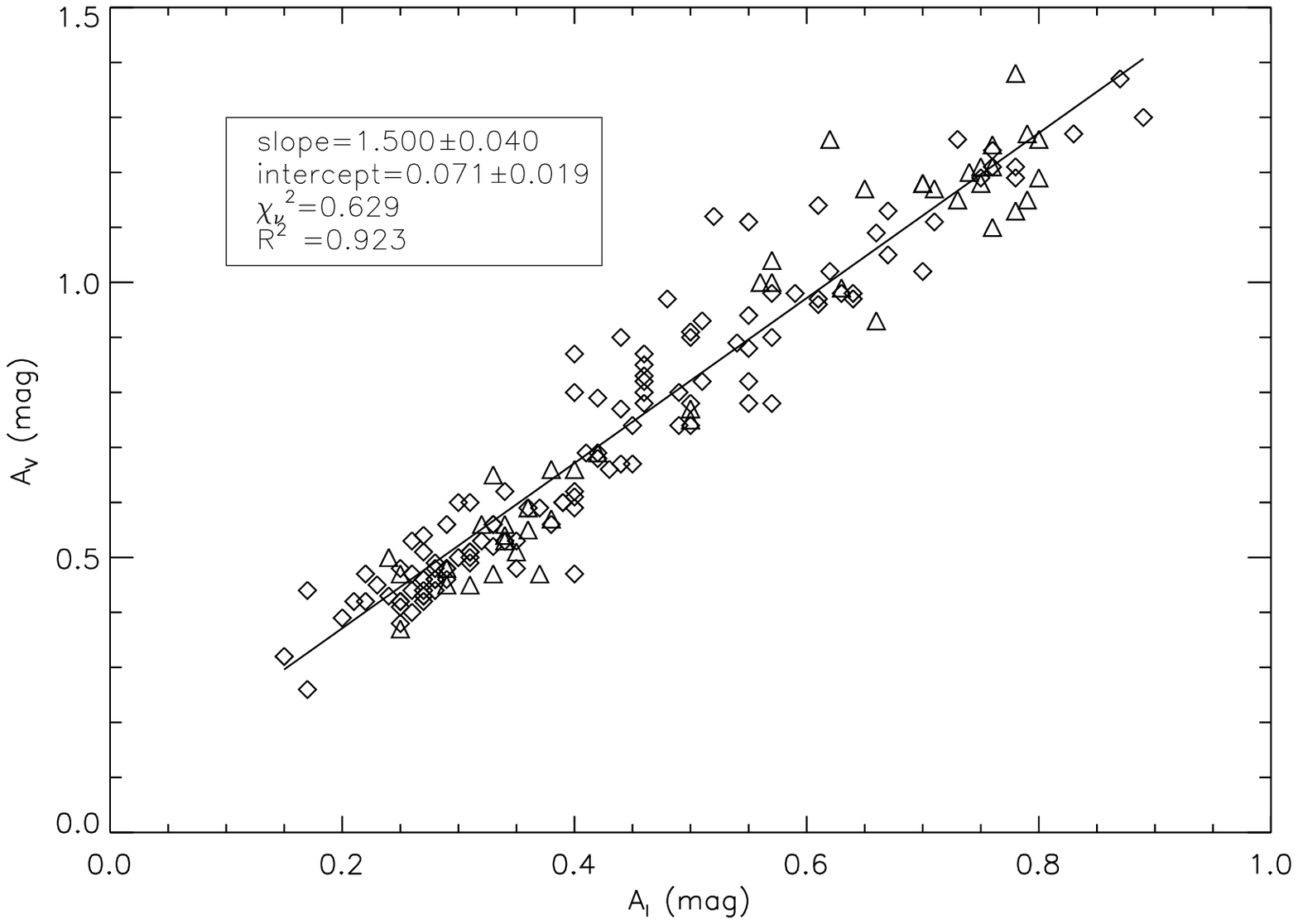}
\caption{V band amplitudes as functions of I band amplitudes for RRL
stars in 4 globular clusters. Solid line denotes the linear relation
as determined by a least squares fit. The diamonds denote the
observational data taken from Walker 1994 (M68), Walker \& Nemec
1996 (IC 4499) and Walker 1998 (NGC 1851). The triangles denote the
amplitude relations determined by us from the published light curves
from Benk\H{o} et al. (2006)  (M3). The quadratic correlation coefficient is
0.923.}
\label{Fig 12}%
\end{figure}

\begin{figure}
\centering
\includegraphics[height=8cm,width=8.5cm]{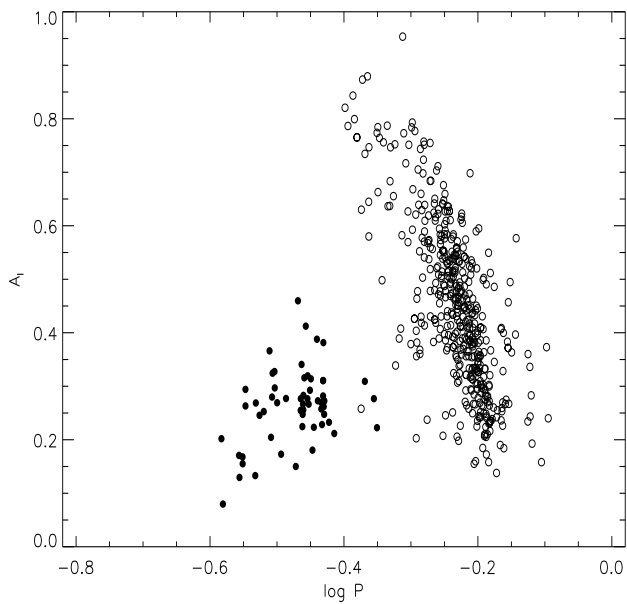}
\caption{Period-amplitude diagram for all RRab (open circles) and
RRc stars (filled circles). }
\label{Fig 13}%
\end{figure}

\subsection{Mean magnitudes in V band}
Calibration to the standard V band magnitude is  necessary for
calculation of the distance scale. The mean magnitude was calculated
for each object from its light curve.  To transform the I band mean
magnitude to the V band we used RRL light curves in the V and I band
from the OGLE database itself. Typical errors in the V band
magnitudes in OGLE are 0.02 - 0.07 mag. For RRab stars a linear
least-square fit resulted in the following conversion relation :
\begin{eqnarray}
\rm A_{0}(\rm V) = -0.691\,(\pm 0.029) + 1.068\,(\pm 0.002)\,\rm A_{0}(\rm I).
\end{eqnarray}
For the RRc stars the result of the linear least square fit is as
follows:
\begin{eqnarray}
\rm A_{0}(\rm V) = 3.733\,(\pm 0.536) + 0.831\,(\pm 0.028)\,\rm A_{0}(\rm I).
\end{eqnarray}
The fits  of Eqns.~(9) \& (10) are  shown in Fig.~14. Light curves
having  mean magnitude variation $\ge 2 \sigma$ were discarded and V
band magnitude thus determined. On the other hand, the absolute
magnitude calculation of RRab stars required \rm A$_{1}$ while RRc stars
required \rm A$_{4}$ in V band. We used V and I band Data from Walker
(1994, 1998) for the \rm A$_{4}$ V band calibration and a linear
regression analysis yielded the following relation
\begin{eqnarray}
\rm A_{1}(\rm V) = -0.007\,(\pm 0.001) + 1.686\,(\pm0.040)\,\rm A_{1}(\rm I),
\end{eqnarray}
for RRab stars and
\begin{eqnarray}
\rm A_{4}(\rm V) = 0.001\,(\pm 0.0007) + 1.056\,(\pm 0.115)\,\rm A_{4}(\rm I),
\end{eqnarray}
for RRc stars. Following Saha \& Hoessel (1990) and  Sakai et al.
(1999), the intensity weighted mean magnitudes and phase-weighted
mean magnitudes are given by
\begin{eqnarray}
\rm m_{\rm int}=-2.5 \rm \log \rm \sum_{\rm i=1}^{\rm n} \rm \frac
{\rm 1} {\rm n} 10^{\rm -0.4\rm m_{\rm i}},
\end{eqnarray}
\begin{eqnarray}
\rm m_{\rm ph}=-2.5 \rm \log \rm  \sum_{\rm i=1}^{\rm n} 0.5(\rm
\phi_{\rm i+1}-\rm \phi_{i-1}) 10^{-0.4\rm m_{\rm i}},
\end{eqnarray}
 \\
where n is the total number of observations, m$_{\rm i}$ and
$\phi_{\rm i}$ are the magnitude and phase of the $i$th observation
respectively, in order of increasing phase. The intensity-weighted
mean magnitude and phase-weighted mean magnitude in the V band have
been determined using a linear least square fit between the V and I
band data resulting in the following conversion relations
\begin{eqnarray}
\rm m_{\rm int}(V) = 0.862\,(\pm0.360) + 0.984\,(\pm0.019)~\rm
m_{\rm int}(I),
\end{eqnarray}
\begin{eqnarray}
\rm m_{\rm ph}(V) = 0.471\,(\pm 0.320) + 1.009\,(\pm0.021)~\rm
m_{\rm ph}(I),
\end{eqnarray}
for RRab stars, and
\begin{eqnarray}
\rm m_{\rm int}(V) = 3.093\,(\pm 0.867) + 0.863\,(\pm 0.045)~\rm
m_{\rm int}(I),
\end{eqnarray}
\begin{eqnarray}
\rm m_{\rm ph}(V) = 5.900\,(\pm 0.893) + 0.733\,(\pm0.045)~\rm
m_{\rm ph}(I),
\end{eqnarray}
for RRc stars.
\begin{figure*}
\centering
\includegraphics[height=13cm,width=17cm]{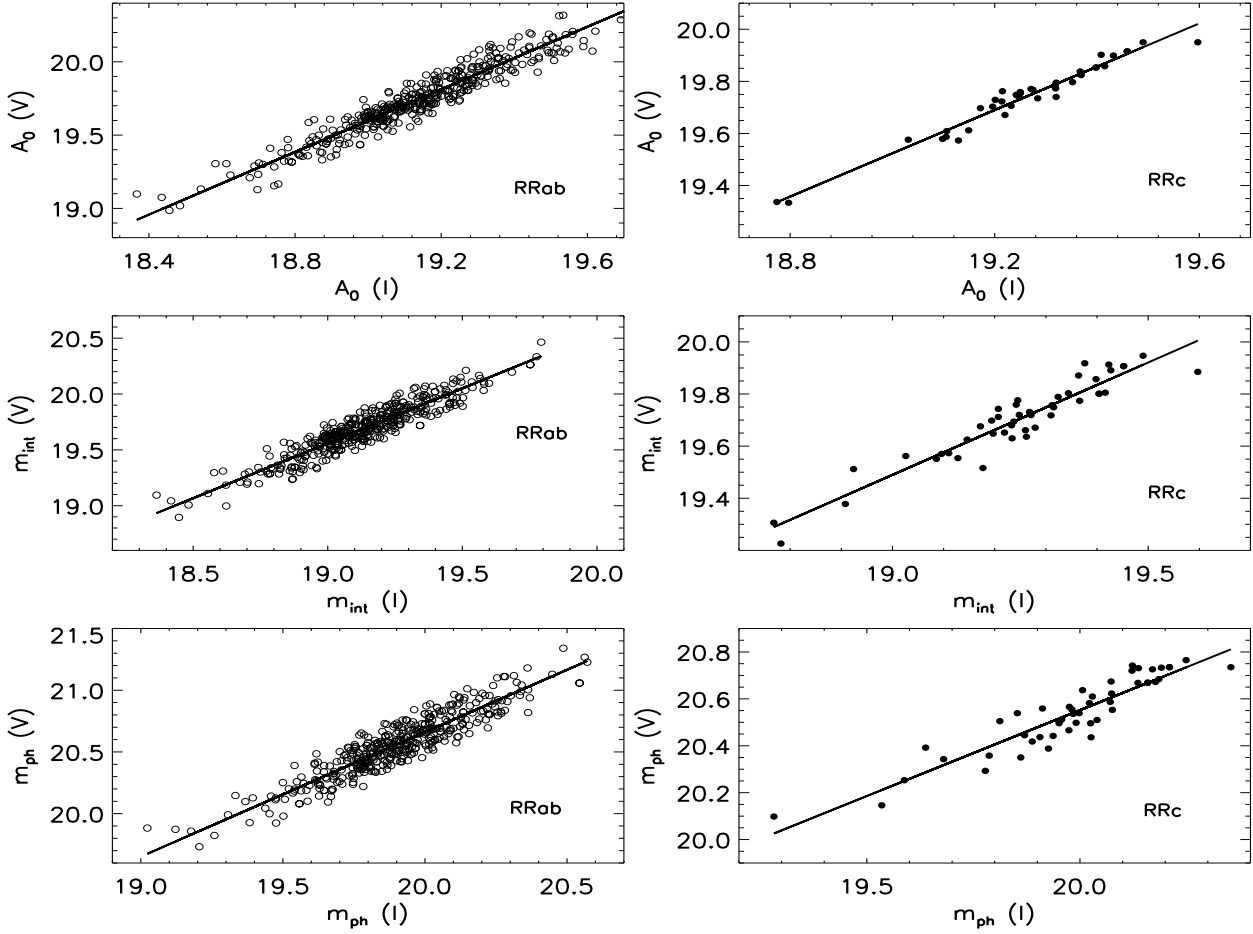}
\caption{Mean magnitude V band calibration for RRab stars (open
circles) and RRc stars (filled circles). Top panels show fits
represented by Eqns.~(9) \& (10). The lower panels show mean
magnitude calibration using both V and I band data available for 465
RRab and 55 RRc stars in SZ02. }
\label{Fig 14}%
\end{figure*}
\section{Physical Parameter extraction}
\subsection{RRab stars}
Using a set of 81 RRab stars, JK96 derived an empirical relation
between the Fourier decomposition parameter ${\phi_{31}}$, period
(P) and metallicity [Fe/H] given by
\begin{equation}
 [\rm Fe/\rm H] = -5.038 - 5.394\, \rm P + 1.345\, \rm \phi_{31}.
\end{equation}
The Fourier parameters in the above relation were obtained for a
sine Fourier series fit. Care has to be taken that the RRab stars to
which this equation is applied follow the light curve systematics
defined by calibrating the sample of JK96. This can be ensured if
the light curve in question satisfies a certain compatibility
criterion. The various parameters of the calibrating data set obey
well-defined correlations (see Table 6 of JK96). Deviation of a
light curve from the calibrating characteristics is quantified via
the deviation parameter D$\rm _F$ defined as
\[\rm D_{\rm F}=\frac
{\rm F_{\rm obs}-\rm F_{\rm calc}}{\rm \sigma_{\rm F}},\] where
F$_{\rm obs}$ is the observed value of a given Fourier parameter,
$\rm \sigma_{F}$ is the corresponding deviations of various
correlations. If the maximum value of \rm D$_{\rm
F}$ of each parameter is less than 3, then each light curve
satisfies the compatibility condition.  This maximum value of
D$_{\rm m}$ represents a quality test on the regular behaviour of
the shape of the light curve. But when the test is applied to the
RRab stars of OGLE SMC RR lyrae stars, some of the stars did not pass this 
criterion even though the shapes of
the light curve are `normal looking'. One of the possible reasons is
that the OGLE SMC RRL light curves are constructed from the
observation in the I band but not in the V band on which the
deviation parameter is defined.  We have noticed that for the deviation
parameter up to a value of \rm D$_{\rm m} = 5$, the shapes of the
RRab light curves are normal. This value of D$_{\rm m}$ is used as a
cutoff limit for the selection of clean samples of RRab stars for
analysis. The deviation parameter of individual RRab light curves of
the OGLE data has been calculated and is given in the last column of
Table 1. Cacciari et al. (2005) have also adopted this slightly
relaxed criterion to improve the statistics after they verified that this 
does not lead to any significant difference in the resulting physical 
parameters. For the RRab analysis, we apply various empirical relations from the
literature to all our target stars only when \rm D$_{\rm m} \le 5$.
After implementing this compatibility test we have found that 355
RRab stars out of 478 are now available for further analysis. This
means that our sample of RRab stars now contains 355 RRab stars that
have `normal looking' light curve shapes. It is also seen that the
maximum contribution to the deviation parameter comes from the deviation in 
$\phi_{31}$.

The intrinsic colors as derived from the Fourier parameters of the RRab
stars can be used to estimate the temperature of these stars.
These color indices as defined by Jurcsik (1998) are the differences of
the magnitude-averaged absolute brightness. The intrinsic
color relations is of the form
\begin{eqnarray}
\rm (B-V) = 0.308 + 0.163 \, \rm P - 0.187 \, \rm A_{1}.
\end{eqnarray}
The above relation can be used to calculate the effective temperature
\begin{eqnarray}
\rm log\, \rm T_{eff}\,\rm (B-V) = 3.930 - 0.322\,\rm (B-V) +
0.007\,[\rm Fe/H].
\end{eqnarray}
The relation between the absolute magnitudes and the Fourier
parameters is given by the following relations
\begin{eqnarray}
\rm M_{V} = 1.221 - 1.396\,\rm P - 0.477\,\rm A_{1} + 0.103\,\rm
\phi_{31}.
\end{eqnarray}
Eqn.~(20) is from Jurcsik (1998) and Eqn.~(21) is taken from
Kov\'{a}cs \& Walker (2001). Eqn.~(22), taken from Kov\'{a}cs (1998)
is based on the relation between the intensity-averaged M$_{\rm V}$ and
the V band Fourier parameters. On the other hand on  a different
absolute scale, Jurcsik (1998) derived a relation for the mass of
the RRab stars as follows
\begin{eqnarray}
\rm log\,\rm (M/M_{\odot}) = 20.884 - 1.754\,\rm \log\,\rm P + 1.477\, ~~~~\nonumber \\
~~\rm log\,(\rm L/L_{\odot})  
- 6.272\,\rm log\,\rm T_{\rm eff}+0.0367\,\rm [\rm Fe/H].
\end{eqnarray}

The Fourier coefficients in the above equations are based on a sine
series fit. But the coefficients in the present analysis (Table 1) are 
calculated for a cosine series fit. Therefore, to put the coefficient 
$\rm \phi_{31}$ on to the appropriate system, we add 3.145 to $\rm \phi_{31}$ 
in Table 1. In Table 5 we list various physical parameters computed from the 
above empirical relations for 335 RRab stars with regular light curves with  
$\rm D_{m} \le 5$ . Mean error in the Fourier phase parameter 
$\phi_{31}$ for 335 RRab stars is $\sim$ 0.19. This should result in an 
uncertainty of $\sim$ 0.25 dex in [Fe/H], $\sim$ 0.02 mag in the absolute 
magnitude, $\sim$ 12 K in temperature, $\sim$ 0.01 M$_{\odot}$ in mass.

\subsection{RRc stars}
For the RRc stars, we have only retained stars for which the error in 
$\phi_{31}$  is less than 0.5. It is generally not desirable to have error 
larger than 0.2 (Kaluzny et al. 1998).
 Since the RRc stars in the SMC are fainter and have lower 
amplitudes, good data for such sources are simply not available at present.
Imposing the criterion of $\sigma_{\phi_{31}} < 0.3$ leaves us with only 3 RRc 
stars to be used for further analysis. To increase the statistics without 
significant loss of accuracy, we have chosen $\sigma_{\phi_{31}} < 0.5$ which 
allows us to analyse 17 RRc stars.

Through Hydrodynamical pulsation models, SC93  formulated the
following relationships for Luminosity, mass and helium abundance
(Y) of RRc stars :
\begin{equation}
\rm log\, (\rm L/L_{\odot}) = 1.04\,\rm \log\,\rm P - 0.058\,\rm \phi_{31} +
2.41,
\end{equation}
\begin{equation}
\rm \log\,(\rm M/M_{\odot}) = 0.52\,\rm \log\,\rm P - 0.11\,\rm \phi_{31} +
0.39,
\end{equation}

\begin{equation}
\rm \log\,\rm (T_{\rm eff}) = 3.7746 - 0.1452\,\rm \log\,\rm P +
0.0056\,\rm \phi_{31}.
\end{equation}                                                  

\begin{eqnarray}
\log\, \rm Y = -20.26 + 4.935\,\rm \log\,\rm (T_{eff}) - 0.2638 ~~~~~~\nonumber \\
\,\rm \log\,\rm (M/M_{\odot}) + 0.3318\,\rm \log\,(\rm L/L_{\odot}).
\end{eqnarray}
The rms error to the fit in Eqn.~(24) is 0.025.

Another alternative to Eqn.~(24) is the expression for the intensity averaged 
absolute magnitude proposed by Kov\'{a}cs (1998, hereafter K98) based on 
observations of RRc variables in eight different stellar systems:
\begin{equation}
\rm M_{\rm V} = 1.261 - 0.961\,\rm P - 0.044\,\rm \phi_{21} - 4.447
\,A_{4},
\end{equation}
with an rms of 0.042. Kov\'{a}cs derived this formula from the light 
curves of 106 RRc stars, mainly from Sculptor and M68 and calibrated the zero 
point from the Baade-Wesselink luminosity scale of Clementini et al. (1995). 
The phase difference $\phi_{21}$ in Eqn. (28) is based on a sine series 
fit. Therefore, 1.571 is subtracted from $\phi_{21}$ values in Table 2, which 
are based on a cosine series fit.

Following L\'{a}zaro et al. (2006), we have calculated the luminosities of 
SMC RRc stars from Eqn.~(24) as well as from Eqn.~(28). The results are 
plotted in Fig.~15. 
Luminosities calculated using SC93 relation (Eqn.~(24)) are larger than those 
estimated by the K98 relation (Eqn.~(28)) and the difference may be attributed
to different scaling (zero points) of the two relations. 
The zero point of the luminosity scale of RRc stars is still 
a matter of controversy (Cacciari et al. 2005, Nemec 2004, L\'{a}zaro et al. 2006). 
Some problems with the Simon's calibration for the RRc stars were also noted 
by Catelan (2004) who concluded that SC93's relations for L and M cannot both 
be simultaneously valid. Eqns. of SC93 (Eqns. (24) \& (25)) yield
\begin{equation}
\rm \log\,P = -2.877+1.305\,\log\,(\rm L/L_{\odot})-0.688\log\,(\rm M/M_{\odot}),
\end{equation}

The above relation lacks a temperature term which should be 
present according to the period-density relation (Catelan 2004). 
This is true and remains an unresolved theoretical issue.

The empirical relation for metallicity in terms of the Fourier
decomposition parameter $\rm \phi_{31}$ was derived by Morgan et al.
(2007) by considering a
large number of stars in different globular clusters which have the
V band Fourier coefficients available in the literature and the
sources of metallicities available for these cluster members. Using
the metallicty scales of Zinn \& West (1984) and Carretta \& Gratton
(1997), they obtained the following two empirical relations for RRc stars
respectively

\begin{eqnarray}
[\rm Fe/\rm H] = 52.466\,\rm P^{2} - 30.075\,\rm P + 0.131\,\rm {\phi_{31}}^2 +~~~~~~\nonumber \\
0.982\,\rm \phi_{31} - 4.198\,\rm P\rm \phi_{31} + 2.424,
\end{eqnarray}
which has a sample standard deviation of 0.145 dex, and
\begin{equation}
[\rm Fe/\rm H] = 0.034\,\rm {\phi_{31}}^2 + 0.196\,\rm \phi_{31}
-8.507\,P + 0.367,
\end{equation}
which has a sample standard deviation of 0.142 dex. We have used Eqn. (31) for 
the calculation of [Fe/H].

The temperatures determined using Eqn. (26) are found to be
overestimated for some of the RRc stars. Cacciari et al. (2005) also
used Eqn.~(26) and found a similar trend for temperature estimates
of RRc stars of globular cluster M3 when compared with temperatures
derived by Sekiguchi \& Fukugita (2000) from the B-V Colors.
Temperature differences as large as 500 K have been found for some
of the RRc stars. Therefore, the temperature of RRc stars as
determined from Fourier coefficients need to be re-examined. 
 The inaccuracy of the empirical relations connecting the effective temperature,
$\rm \phi_{31}$ term and the period is also reflected in the calculation
of the radius of RRc stars. We show ahead that the radii of RRc
stars determined from the Fourier decomposition technique differ 
largely from those determined from the theoretical
period-radius-metallicity relation of Marconi et al. (2005). On the
other hand, for RRab stars there is a well-matched similarity
between the radii determined from the Fourier decomposition
technique and from the more fundamental relations used by Marconi et
al. (2005). Therefore, the empirical relations derived for the RRab stars 
seem to be accurate, whereas the physical parameter estimation using the
empirical relations derived for the RRc stars may be unreliable. In Table 6, we list the various physical parameters computed from the above 
empirical relations for the RRc stars with $\sigma_{\phi_{31}} < 0.5$.

Mean error in the Fourier phase parameter 
$\phi_{31}$ for 17 RRc stars is $\sim$ 0.396. This will result in an 
uncertainty of $\sim$ 0.14 dex in [Fe/H],  $\sim$ $0.02$ in $\log\,(\rm 
L/L_{\odot})$, $\sim$ 16 K in temperature and $\sim$ 0.18 M$_{\odot}$ in mass.

\subsection{HR Diagram}
An H-R diagram of the SMC RRLs stars is plotted in Fig.~15
along with the theoretical blue and red edges of the instability
strip. The RRc stars are represented by solid circles whereas the
RRab stars are represented by open circles. Also shown are the
blue and red edges of the instability strip of Fundamental (RRab)
and first-overtone (RRc) variables derived from the theoretical
convective pulsation models of Bono, Caputo \& Marconi (1995). It
can be easily seen that all the RRab and RRc stars lie well within
their instability strips. Although the RRc stars lie inside
the instability strip, the empirical relations overestimate the temperatures 
of these stars.
\begin{figure}
\centering
\includegraphics[height=8cm,width=8.5cm]{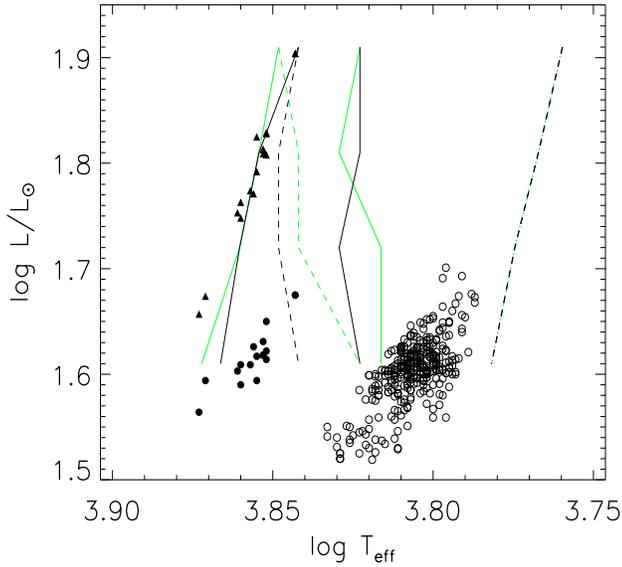}
\vspace{10pt} \caption{H-R diagram of 335 RRab and 17 RRc stars and the 
theoretical pulsational blue and red edges of the instability strip from 
Bono et al. (1995).  Open circles represent RRab stars, solid circles represent
RRc stars with $\log\,(\rm L/L_{\odot})$ calculated from Eqn. (28) and filled upper 
triangles RRc stars with $\log\,(\rm L/L_{\odot})$ calculated from Eqn. (24).  
The green solid and dashed lines show the boundaries of the instability
strip for RRc and RRab stars respectively for mass 0.75M$_{\odot}$
and the black solid and dashed lines show the boundaries of the
instability strip for RRc and RRab stars respectively for mass
0.65M$_{\odot}$.}
\label{Fig 15}%
\end{figure}
\subsection{Gravity for RRab and RRc stars}
From the global physical parameters like mass, luminosity and
temperature, the gravity can be calculated from the following
equation (cf. Cacciari et al. 2005)
\begin{equation}
\rm \log{\rm  g}=-10.607+\rm \log\,(\rm M/M_{\odot})-\rm \log\,(\rm L/L_{\odot})+4\rm \log\rm T_{\rm eff}
\end{equation}
\begin{figure}
\centering
\includegraphics[height=10cm,width=8cm]{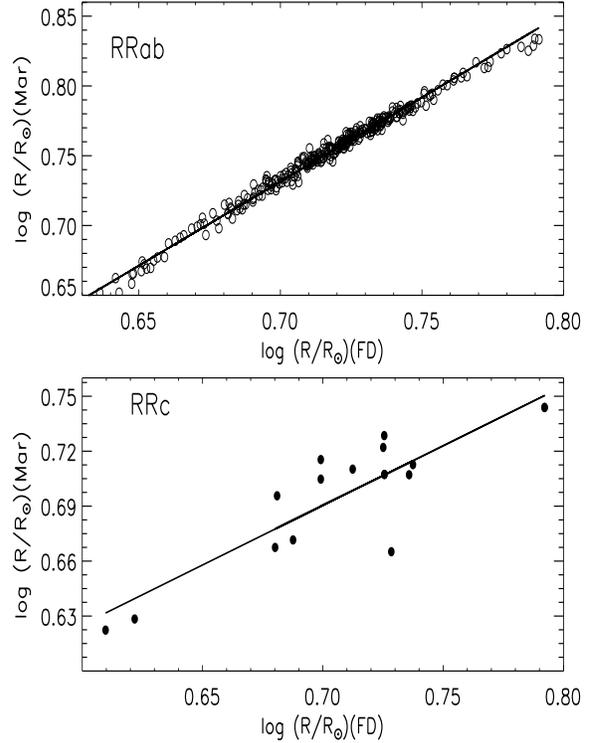}
\caption{Radii obtained from the empirical relation of Marconi et
al. (2005) are plotted versus the radii determined from the Fourier
decomposition (FD) method. For RRab (open circles) stars in the
upper panel the linear regression analysis yields : log\,
(R/R$_{\odot}$)\,(Mar)\,=\,-0.112\,($\pm0.005$)\,+\,1.206($\pm0.007$)\,log
(\,R/R$_{\odot}$)\,(FD) and for the RRc stars (filled circles) in
the lower panel the linear regression analysis yields: log
\,(R/R$_{\odot}$)\,(Mar)\,= 0.235($\pm0.064$)\,+\,0.651($\pm0.090$)\,log
(\,R/R$_{\odot}$) \,(FD).}
\label{Fig 16}%
\end{figure}
\subsection{Radii of RRLs}
The radii of the RRL can be obtained once we know the
temperature and the luminosity of the RRLs stars using the
Stefan-Boltzmann law. 
Marconi et al. (2005) have recently given a new theoretical
period-radius-metallicity relation for the RR
 Lyraes  based on detailed and homogeneous set of nonlinear models with
 a wide range of stellar masses and chemical compositions.
 We show how the radii obtained by the Fourier
decomposition method compare to the radii obtained from the
period-radius-luminosity (PRZ) relation of Marconi et al. (2005).
For the sake of completeness, their PRZ relations are given as follows:
\begin{eqnarray}
\log\,\rm R = 0.774(\pm 0.009) + 0.580(\pm 0.007)\,\log\,\rm P ~~~~~ \nonumber \\
- 0.035(\pm0.001)\,\log \,\rm Z,
\end{eqnarray}
for RRab stars with $\sigma = 0.008$.
\begin{eqnarray}
\log \,\rm R = 0.883(\pm 0.004) + 0.621(\pm 0.004)\, \log\,\rm P
~~~~~~~\nonumber \\
- 0.0302(\pm 0.001)\,\log \,\rm Z,
\end{eqnarray}
for RRc stars with $\sigma = 0.004$. $\log\,\rm Z$ can be calculated
from the relation
\begin{eqnarray}
\log \,\rm Z = [\rm Fe/H] - 1.70 + \log \, (0.638\,\rm f + 0.362),
\end{eqnarray}
where $\rm f$ is an $\alpha$-enhancement factor with respect to iron
(Salaris, Chieffi \& Straniero 1993). We take $\rm f=1$. In Fig.~16, we
compare our determinations of the radii using the Fourier
decomposition technique (FD) with those obtained by Marconi et al.
(2005) empirical relations (Mar). 
In case of the radius determinations of RRc stars by the FD method, the 
SC93's equations for $\log\,\rm L$ and $\log\,\rm T$ are used. It can be seen that the 
radii obtained by the two different methods are quite consistent for RRab stars.
The solid line in Fig.~16 represents the best least square fit. The correlation
coefficient between the two radii determinations for the RRab stars are 0.994, whereas for the RRc stars (lower panel of Fig.~16), the correlation coefficient
 is 0.880. This suggests that the radii determinations by the two independent 
methods are strongly correlated for the RRab stars, whereas this is not the 
case for the RRc stars.

\section {Distance to the SMC}

We use mean magnitude A$_{0}$(V), intensity-weighted mean magnitude
and phase weighted mean magnitude separately to derive an
independent  distance modulus to the SMC. The distance modulus for
each star is calculated from their values of mean magnitudes and
M$_{V}$ derived from the Fourier parameters. Assuming the reddening
estimates of the eleven SMC fields of SZ02 to be accurate, we take
average of the reddening values of these fields as the true
estimates of E(B-V). Following SZ02 we take the interstellar
extinction as A$_{V}$=3.24\,E(B-V).  This value of A$_{V}$ has been
used to estimate the  reddening free distance modulus of SMC. Using
335 RRab and 17 RRc stars  we find the mean distance modulii  of
SMC to be 18.89$\pm$0.01 mag, 18.87$\pm$ 0.03 mag and 18.87
$\pm 0.02$ mag from the mean magnitude, intensity-weighted mean
magnitude and phase-weighted mean magnitude respectively for all the
352 RRL stars. The uncertainty is the standard deviation of the
mean from the individual stars. The mean distance modulus so
determined are consistent with the earlier values of 19.05$\pm$
0.017 (Kov\'{a}cs et al. 2000), 18.89$\pm$0.4 (Harries et al. 2003)
and 18.91$\pm$0.1 (Hilditch et al. 2005).
\section{Summary \& Conclusions}
In this paper we have derived the physical parameters of 352 RR
Lyrae stars (335 RRab and 17 RRc) of the SMC from OGLE-II I band
database using the Fourier decomposition of their light curves. The
stars were selected based on the quality of their light curves. I
band Fourier coefficients have been converted to V band using the
inter-relations obtained from the observational data of B06 and  
W94. Using the Fourier decomposition
method, we find the mean physical parameters: 
[Fe/H] = -1.56\,$\pm\,0.25$, M = 0.55
$\pm $\,0.01\,M$_{\odot}$, T$_{\rm eff} = 6404\, \pm 12$ K,
$\log \rm L = 1.60\,\pm0.01\,\rm L_\odot$ and M$\rm _V = 0.78\,\pm0.02 $ 
for 335 RRab variables and [Fe/H] = -1.90\,$\pm$\, 0.13, M = 0.82
$\pm $\,0.18\,M$_{\odot}$, T$_{\rm eff} = 7177\,\pm 16$ K, $\log\,\rm L = 1.62\,\pm \,0.02\,\rm L_{\odot}$ and M$\rm _V = 0.76\,\pm 0.05$ for 17 RRc stars. Mean distance modulus to the SMC was calculated from the 352 light 
curves by using mean magnitude, intensity weighted mean magnitude and phase 
weighted mean magnitude. The values of the distance modulus are found to be in
good agreement with independent studies.  Locations of the RRL
stars in the H-R diagram clearly show that the estimates of the
parameters determined from the Fourier decomposition method are
consistent with the theoretical blue and red edges of the
instability strip calculated by Bono et al. (1995). The calculations of the radii of the RRLs by using the Fourier decomposition technique are in agreement 
with the theoretical period-radius-metallicity relations of Marconi et al. 
(2005) obtained from a completely different approach of non-linear convective 
modelling. 
\begin{figure}
\centering
\includegraphics[height=8cm,width=9cm]{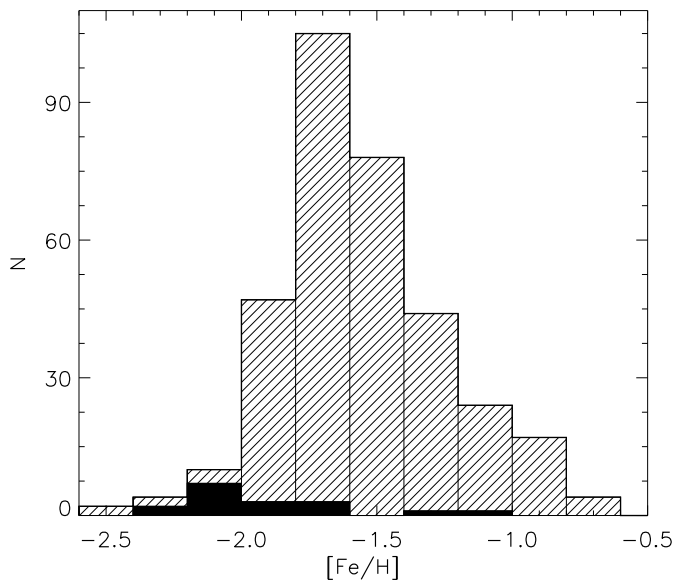}
\vspace{10pt} 
 \caption{Histogram of the metallicity of the RRLs used for the 
analysis. The partially filled histogram shows the metallicity distribution 
of the 335 RRab stars, whereas the fully filled histogram shows the metallicity
distribution of the 17 RRc stars.}
\label{Fig 14}%
\end{figure}

\begin{table*}
\caption{Physical parameters extracted from Fourier coefficients for
 335 RRab variables. Errors represent the uncertainties in the Fourier 
parameters.}
\begin{center}
\scalebox{0.9}{
\begin{tabular}{ccccccccccccccccc}
\\
\hline \hline
OGLE ID &M$_{V}$&\rm $\log$\,(L/L$_{\odot})$&[\rm Fe/H]&T$_{\rm eff}$&\rm M/M$_{\odot}$&\rm log{\rm  g}&$\rm R/R_{\odot} (FD)$& $\rm R/R_{\odot} $ (Mar)  \\
~~~~~~~(1)&(2)&(3)&(4)&(5)&(6)&(7)&(8)&(9)\\

\hline
\hline
 004801.59-733021.5&   0.93$\pm$   0.02&    1.53$\pm$   0.01&   -0.91$\pm$   0.19&  6751$\pm$ 9&    0.61$\pm$   0.01&    2.97$\pm$   0.02&    4.27$\pm$   0.02&    4.31$\pm$   0.03\\
  005300.26-725136.6&   0.92$\pm$   0.02&    1.54$\pm$   0.01&   -1.18$\pm$   0.19&  6702$\pm$ 9&    0.64$\pm$   0.01&    2.97$\pm$   0.02&    4.40$\pm$   0.02&    4.43$\pm$   0.03\\
  003841.21-734422.9&   0.90$\pm$   0.01&    1.54$\pm$   0.01&   -1.00$\pm$   0.18&  6758$\pm$ 9&    0.60$\pm$   0.01&    2.95$\pm$   0.02&    4.33$\pm$   0.02&    4.41$\pm$   0.03\\
  003727.78-731454.9&   0.92$\pm$   0.01&    1.53$\pm$   0.01&   -0.88$\pm$   0.16&  6755$\pm$ 8&    0.59$\pm$   0.01&    2.95$\pm$   0.01&    4.29$\pm$   0.02&    4.38$\pm$   0.02\\
  005728.85-723454.6&   0.93$\pm$   0.02&    1.52$\pm$   0.01&   -0.76$\pm$   0.21&  6739$\pm$10&    0.57$\pm$   0.01&    2.94$\pm$   0.02&    4.26$\pm$   0.02&    4.36$\pm$   0.03\\
  005728.85-723454.6&   0.93$\pm$   0.02&    1.52$\pm$   0.01&   -0.76$\pm$   0.21&  6739$\pm$10&    0.57$\pm$   0.01&    2.94$\pm$   0.02&    4.26$\pm$   0.02&    4.36$\pm$   0.03\\
  005026.32-732418.2&   0.94$\pm$   0.02&    1.52$\pm$   0.01&   -1.00$\pm$   0.20&  6653$\pm$ 9&    0.60$\pm$   0.01&    2.94$\pm$   0.02&    4.39$\pm$   0.02&    4.48$\pm$   0.03\\
  004639.18-731324.7&   0.89$\pm$   0.02&    1.54$\pm$   0.01&   -0.81$\pm$   0.20&  6800$\pm$10&    0.56$\pm$   0.01&    2.93$\pm$   0.02&    4.29$\pm$   0.02&    4.42$\pm$   0.03\\
  003816.46-732449.2&   0.92$\pm$   0.02&    1.52$\pm$   0.01&   -0.74$\pm$   0.19&  6738$\pm$ 9&    0.55$\pm$   0.01&    2.93$\pm$   0.02&    4.29$\pm$   0.02&    4.42$\pm$   0.03\\
  005110.48-730750.0&   0.87$\pm$   0.01&    1.55$\pm$   0.01&   -0.85$\pm$   0.18&  6805$\pm$ 9&    0.55$\pm$   0.01&    2.92$\pm$   0.02&    4.33$\pm$   0.02&    4.48$\pm$   0.03\\
  010452.90-724025.9&   0.95$\pm$   0.01&    1.52$\pm$   0.01&   -1.00$\pm$   0.19&  6599$\pm$ 9&    0.59$\pm$   0.01&    2.92$\pm$   0.02&    4.44$\pm$   0.02&    4.55$\pm$   0.03\\
  005646.16-723452.2&   0.88$\pm$   0.01&    1.55$\pm$   0.01&   -1.06$\pm$   0.18&  6691$\pm$ 9&    0.57$\pm$   0.01&    2.90$\pm$   0.02&    4.48$\pm$   0.02&    4.65$\pm$   0.03\\
  005957.83-730647.6&   0.87$\pm$   0.01&    1.55$\pm$   0.01&   -0.99$\pm$   0.17&  6717$\pm$ 8&    0.56$\pm$   0.01&    2.90$\pm$   0.01&    4.44$\pm$   0.02&    4.63$\pm$   0.03\\
  005458.09-724948.9&   0.89$\pm$   0.01&    1.55$\pm$   0.00&   -1.11$\pm$   0.15&  6646$\pm$ 7&    0.58$\pm$   0.01&    2.90$\pm$   0.01&    4.51$\pm$   0.01&    4.68$\pm$   0.02\\
  004758.98-732241.3&   0.89$\pm$   0.02&    1.54$\pm$   0.01&   -0.95$\pm$   0.26&  6683$\pm$12&    0.56$\pm$   0.01&    2.90$\pm$   0.02&    4.45$\pm$   0.03&    4.63$\pm$   0.04\\
  010535.93-720621.6&   0.94$\pm$   0.02&    1.53$\pm$   0.01&   -0.98$\pm$   0.30&  6583$\pm$13&    0.57$\pm$   0.01&    2.89$\pm$   0.03&    4.49$\pm$   0.03&    4.67$\pm$   0.05\\
  010516.55-722526.5&   0.90$\pm$   0.02&    1.53$\pm$   0.01&   -0.76$\pm$   0.20&  6702$\pm$ 9&    0.53$\pm$   0.01&    2.89$\pm$   0.02&    4.38$\pm$   0.02&    4.59$\pm$   0.03\\
  004721.26-731135.5&   0.89$\pm$   0.01&    1.55$\pm$   0.01&   -1.02$\pm$   0.17&  6640$\pm$ 8&    0.56$\pm$   0.01&    2.88$\pm$   0.02&    4.52$\pm$   0.02&    4.73$\pm$   0.03\\
  005504.67-731106.4&   0.91$\pm$   0.01&    1.54$\pm$   0.01&   -1.08$\pm$   0.19&  6583$\pm$ 9&    0.57$\pm$   0.01&    2.88$\pm$   0.02&    4.56$\pm$   0.02&    4.76$\pm$   0.03\\
  004306.71-733527.9&   0.92$\pm$   0.02&    1.53$\pm$   0.01&   -0.89$\pm$   0.20&  6605$\pm$10&    0.54$\pm$   0.01&    2.88$\pm$   0.02&    4.49$\pm$   0.02&    4.70$\pm$   0.03\\

\hline \hline
\end{tabular}
}
\end{center}
Complete table is available in the electronic form.
\end{table*}

\begin{table*}
\caption{Physical parameters extracted from Fourier coefficients for
 17 RRc variables. Errors represent the uncertainties in the Fourier parameters.  }
\scalebox{0.9}{
\begin{tabular}{cccccccccc}
\\
\hline
\hline
OGLE ID &\rm M$_{\rm V}$(K98)&$\rm \log (\rm L/L_{\odot}) (SC93) $&[\rm Fe/H]&T$_{\rm eff}$&\rm M/M$_{\odot}$&\rm log \rm g&$\rm R/R_{\odot} (FD)$&$\rm R/R_{\odot}$ (Mar)  \\
~~~~~~~(1)&(2)&(3)&(4)&(5)&(6)&(7)&(8)&(9) \\
\hline
\hline
 005451.72-723850.4&   0.85$\pm$   0.03&    1.66$\pm$   0.02&   -1.10$\pm$   0.17& 7457$\pm$17&    0.59$\pm$   0.19&    3.00$\pm$   0.33&    4.07$\pm$   0.03&    4.19$\pm$   0.02\\
  005115.64-724739.2&   0.78$\pm$   0.04&    1.67$\pm$   0.02&   -1.21$\pm$   0.16& 7425$\pm$17&    0.62$\pm$   0.19&    3.00$\pm$   0.31&    4.19$\pm$   0.04&    4.25$\pm$   0.02\\
  010245.99-721132.7&   0.83$\pm$   0.04&    1.83$\pm$   0.03&   -2.02$\pm$   0.11& 7165$\pm$18&    1.13$\pm$   0.07&    3.04$\pm$   0.06&    5.35$\pm$   0.05&    4.63$\pm$   0.02\\
  004631.02-724658.8&   0.78$\pm$   0.04&    1.75$\pm$   0.03&   -1.67$\pm$   0.17& 7266$\pm$20&    0.77$\pm$   0.18&    2.97$\pm$   0.24&    4.79$\pm$   0.05&    4.65$\pm$   0.02\\
  004902.13-724513.6&   0.82$\pm$   0.03&    1.76$\pm$   0.02&   -1.73$\pm$   0.13& 7247$\pm$15&    0.80$\pm$   0.18&    2.97$\pm$   0.23&    4.87$\pm$   0.04&    4.69$\pm$   0.02\\
  010453.59-723756.0&   0.78$\pm$   0.03&    1.83$\pm$   0.01&   -2.11$\pm$   0.08& 7115$\pm$10&    0.93$\pm$   0.17&    2.94$\pm$   0.18&    5.44$\pm$   0.04&    5.09$\pm$   0.01\\
  010029.57-725454.2&   0.77$\pm$   0.04&    1.81$\pm$   0.02&   -2.03$\pm$   0.09& 7136$\pm$10&    0.85$\pm$   0.19&    2.93$\pm$   0.22&    5.32$\pm$   0.05&    5.10$\pm$   0.01\\
  010029.57-725454.2&   0.77$\pm$   0.04&    1.81$\pm$   0.02&   -2.03$\pm$   0.09& 7136$\pm$10&    0.85$\pm$   0.19&    2.93$\pm$   0.22&    5.32$\pm$   0.05&    5.10$\pm$   0.01\\
  004327.25-724542.4&   0.76$\pm$   0.03&    1.75$\pm$   0.02&   -1.62$\pm$   0.13& 7238$\pm$13&    0.64$\pm$   0.22&    2.89$\pm$   0.34&    4.80$\pm$   0.04&    4.96$\pm$   0.02\\
  010231.17-722134.0&   0.69$\pm$   0.04&    1.83$\pm$   0.03&   -2.13$\pm$   0.15& 7107$\pm$18&    0.91$\pm$   0.18&    2.93$\pm$   0.19&    5.46$\pm$   0.06&    5.16$\pm$   0.02\\
  005556.74-732133.6&   0.77$\pm$   0.05&    1.77$\pm$   0.03&   -1.80$\pm$   0.17& 7193$\pm$18&    0.70$\pm$   0.21&    2.89$\pm$   0.31&    5.00$\pm$   0.06&    5.07$\pm$   0.03\\
  003934.35-730433.9&   0.76$\pm$   0.03&    1.79$\pm$   0.02&   -1.92$\pm$   0.13& 7161$\pm$13&    0.76$\pm$   0.21&    2.90$\pm$   0.28&    5.16$\pm$   0.04&    5.13$\pm$   0.02\\
  010647.29-723053.7&   0.73$\pm$   0.04&    1.81$\pm$   0.02&   -2.03$\pm$   0.13& 7124$\pm$14&    0.77$\pm$   0.21&    2.88$\pm$   0.27&    5.31$\pm$   0.04&    5.27$\pm$   0.02\\
  005155.37-724909.5&   0.73$\pm$   0.04&    1.77$\pm$   0.02&   -1.78$\pm$   0.17& 7183$\pm$16&    0.65$\pm$   0.23&    2.86$\pm$   0.35&    5.00$\pm$   0.05&    5.19$\pm$   0.03\\
  010316.37-724816.2&   0.66$\pm$   0.12&    1.90$\pm$   0.03&   -2.56$\pm$   0.13& 6968$\pm$18&    1.14$\pm$   0.12&    2.92$\pm$   0.11&    6.20$\pm$   0.15&    5.54$\pm$   0.02\\
  010316.37-724816.2&   0.66$\pm$   0.12&    1.90$\pm$   0.03&   -2.56$\pm$   0.13& 6968$\pm$18&    1.14$\pm$   0.12&    2.92$\pm$   0.11&    6.20$\pm$   0.15&    5.54$\pm$   0.02\\
  005527.97-724136.8&   0.76$\pm$   0.04&    1.81$\pm$   0.03&   -2.03$\pm$   0.16& 7118$\pm$17&    0.75$\pm$   0.22&    2.87$\pm$   0.29&    5.32$\pm$   0.05&    5.35$\pm$   0.03\\
\hline
\hline
\end{tabular}
}
\end{table*}

\section*{Acknowledgments}
SD thanks CSIR, Govt. of India for a Senior Research Fellowship. HPS
is grateful to CRAL-Observatoire de Lyon for an Invited Professorship.
Authors thank the reviewer Dr. Horace Smith for many useful comments and 
suggestions which improved the presentation of the paper.

\end{document}